\documentclass[graybox]{svmult}

\usepackage{mathptmx}       
\usepackage{helvet}         
\usepackage{courier}        
\usepackage{amsmath}
\usepackage{amssymb}
\usepackage{bm}
\usepackage{type1cm}        
\usepackage{makeidx}         
\usepackage{graphicx}        
\usepackage{multicol}        
\usepackage[bottom]{footmisc}
\usepackage{hyperref}        
\usepackage{soul}            
\usepackage[normalem]{ulem}
\usepackage{multirow}         
\hypersetup{colorlinks=true,urlcolor=blue}
\makeindex             

\newcommand{\ISCO}{{\mbox{\tiny ISCO}}}
\newcommand{\EdGB}{{\mbox{\tiny EdGB}}}

\newcommand{\GR}{{\mbox{\tiny GR}}}

\newcommand{\dCS}{{\mbox{\tiny dCS}}}
\newcommand{\MR}{{\mbox{\tiny MR}}}
\newcommand{\GW}{{\mbox{\tiny GW}}}
\newcommand{\EM}{{\mbox{\tiny GW}}}

\newcommand{\EDGB}{{\mbox{\tiny EdGB}}}
\newcommand{\CS}{{\mbox{\tiny dCS}}}

\definecolor{red(ncs)}{rgb}{0.77, 0.01, 0.2}

\begin{document}
\graphicspath{{./images/}}
\title*{Testing General Relativity with Gravitational Waves}
\author{Zack Carson and Kent Yagi \thanks{corresponding author}}
 \institute{
 Zack Carson \at  Department of Physics, University of Virginia, Charlottesville, Virginia 22904, USA \email{zack.carson@virginia.edu}
 \and
Kent Yagi \at  Department of Physics, University of Virginia, Charlottesville, Virginia 22904, USA, \email{ky5t@virginia.edu}}
%
%
\maketitle

\abstract{
Gravitational-wave sources offer us unique testbeds for probing strong-field, dynamical and nonlinear aspects of gravity. In this chapter, we give a brief overview of the current status and future prospects of testing General Relativity with gravitational waves. In particular, we focus on three theory-agnostic tests (parameterized tests, inspiral-merger-ringdown consistency tests, and gravitational-wave propagation tests) and explain how one can apply such tests to example modified theories of gravity. We conclude by giving some open questions that need to be resolved to carry out more accurate tests of gravity with gravitational waves.
}

\section*{Keywords} 

Gravitational waves; Tests of General Relativity; Black holes; Neutron Stars

\let\cleardoublepage\clearpage

\tableofcontents

\section{Introduction}

The famous observation of gravitational waves (GWs) radiated outwards from the merger of two black holes (BHs) 1.3 billion lightyears away by the LIGO and Virgo Collaborations (LVC) has ushered in the birth of an entirely new era of astrophysics.
For the first time ever, this discovery has allowed us to probe the extreme gravity regime where spacetime is extremely strong, non-linear, and dynamical.
GWs 
carry with them a multitude of fascinating information, including the astrophysical properties of the source BHs, the underlying theory of gravity driving the collision and radiation process, and many more. 
To date, this first GW event and the following $\sim 10$ have failed to present any evidence of deviation from Einstein's famous theory of gravity, general relativity (GR).
This prevailing theory of gravity has remained as its post as the accepted model for the last century since its prescription by Einstein, and has passed every test thrown at it since.
However, a very limited number of tests have been studied in the extreme gravity regimes of spacetimes - such as those surrounding binary BH mergers as detected by the LVC.
Thus, it remains vitally important that we continue to test GR - especially in the unexplored regions of the phase space.

While the current LVC infrastructure is certainly a marvel of modern engineering, it may still not be enough to uncover the elusive traces of a modified theory of gravity, still hiding deep within the current levels of relatively large noise. 
On the other hand, the next generation of GW detectors promise entirely new sensitivity in the mHz regime, as well as incredible improvements on the order of 100 times the sensitivity of current detectors.
As the famous Popper once said, we can never truly prove scientific theories (such as GR), however we can rule out and constrain alternatives with observations~\cite{popper}.
Will these new and improved GW detectors allow us to constrain alternative theories to the point where we begin to see deviations from Einstein's GR?

\begin{figure}[h]
\begin{center}
\includegraphics[width=0.85\columnwidth]{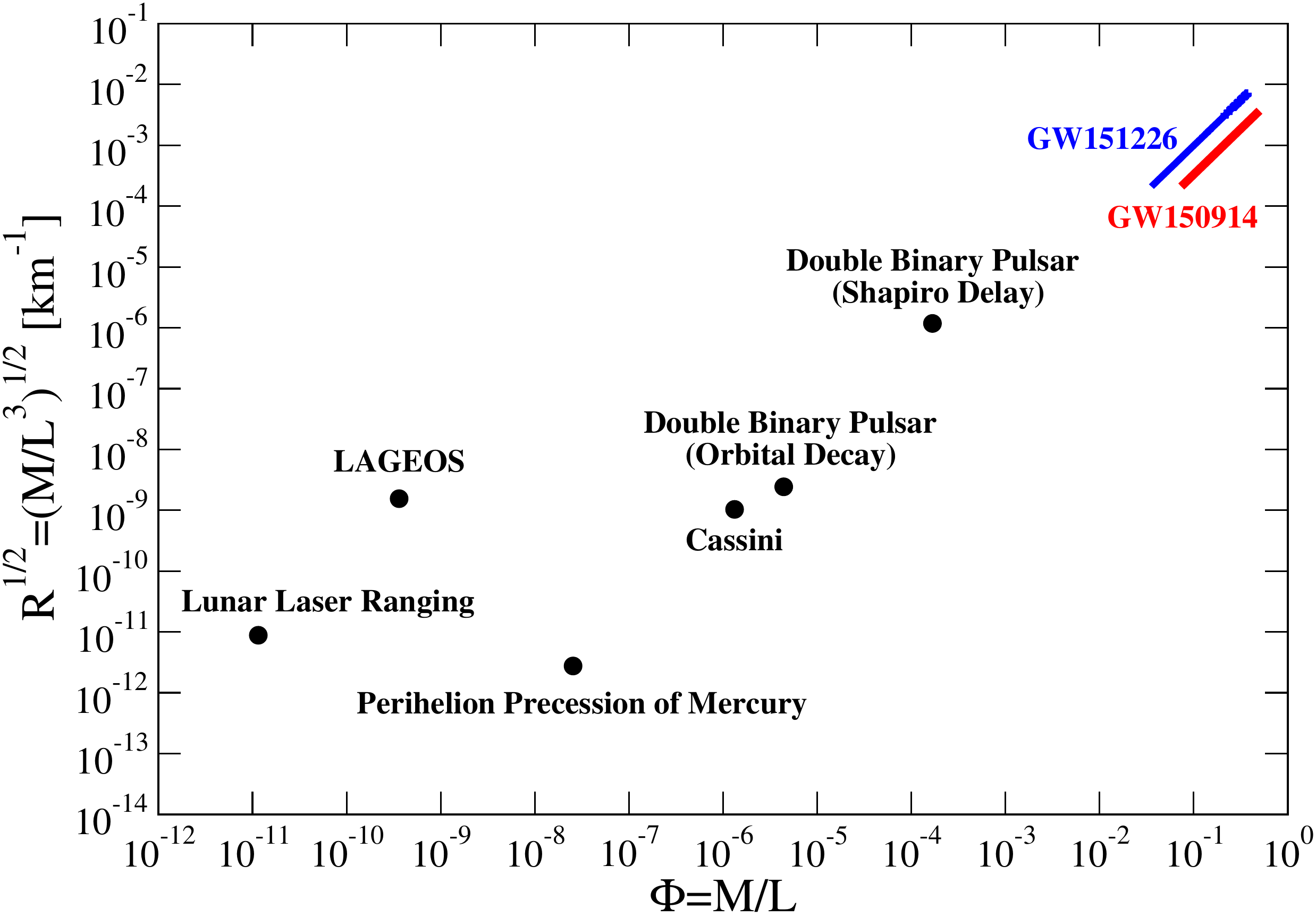}
\end{center}
\caption{Gravitational potential $\Phi$ and (square root of) curvature $\sqrt{R}$ for systems with mass $M$ and size $L$ that have been used to probe GR. The two GW sources (GW150914 and GW151226) lie on the top right corner, meaning that they are strong-field sources. Moreover, these are represented by lines instead of points, which indicates that these sources are also dynamical. 
This figure is taken and edited from~\cite{Yunes_ModifiedPhysics}.
}\label{fig:phase_diagram}
\end{figure}

Gravitational-wave sources have unique features compared to other systems that have been used to probe gravity as they are strong-field, dynamical and nonlinear sources~\cite{Yunes_ModifiedPhysics}. To illustrate this point, we present in Fig.~\ref{fig:phase_diagram} the gravitational potential and curvature of such systems. Notice that the two GW sources have a distinct feature that they lie on the top right corner (strong-field). Moreover, they are shown by lines, which means that they swept through a wide range of potential and curvature during the observational period ($\sim 0.1 - 1$ s), which indicates that they are also dynamical sources (unlike others which are mostly indicated by points). These GW observations can be used to probe various fundamental aspects of GR. These fundamental pillars include the equivalence principle, Lorentz invariance, parity invariance, four-dimensional spacetime, massless gravitons and coordinate commutativity~\cite{Abbott_IMRcon2,Yunes_ModifiedPhysics,Monitor:2017mdv,Abbott:2018lct}. 

\begin{figure}
\begin{center}
\includegraphics[width=0.9\columnwidth]{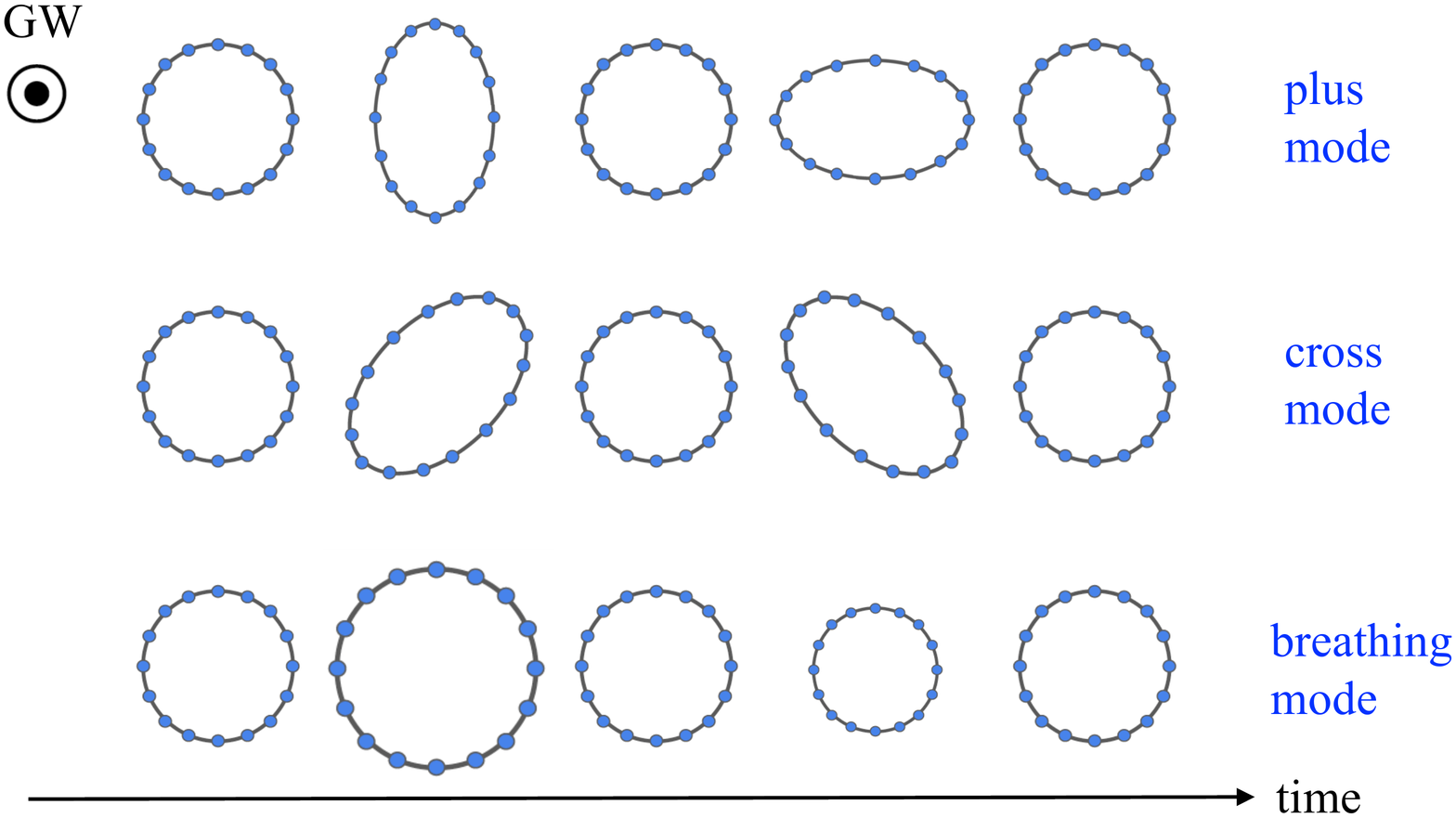}
\end{center}
\caption{Various GW polarizations. This figure shows how a ring of particles move under each GW mode propagating orthogonal to it. GR only contains the plus and cross modes, while in theories beyond GR, there can be additional polarization modes, such as the breathing mode presented here. The first two are tensor modes, while the last one is the scalar mode.
}\label{fig:polarization}
\end{figure}

Another interesting aspect one can probe are GW polarizations. In GR, there are only two tensor polarization modes, which are typically decomposed as plus and cross modes (see Fig.~\ref{fig:polarization}). In theories beyond GR, there can be also two scalar polarization modes (breathing and longitudinal) and two vector modes. An example of the breathing mode is shown in Fig.~\ref{fig:polarization}. The LVC has used GW170817 and carried out a model selection analysis among 3 different models: (i) tensor only, (ii) vector only, and (iii) scalar only models. They found that the first model was significantly preferred from the data over the other two models, and hence GR is consistent with such a test~\cite{Abbott:2018lct}. KAGRA has recently joined the collaboration and started its operation. This additional detector will help to further test additional polarizations. For example, one will be able to compare the tensor-only model against tensor + scalar model which may be more realistic.

In this article, we focus on reviewing three theory-agnostic tests of GR with GWs. The first one is the parameterized test where we add parameterized deviations from GR in the waveform. In certain modified theories of gravity, there are known mappings between such parameters and theoretical constants, and thus the parameterized tests can easily be applied to specific theories. The second test is the inspiral-merger-ringdown (IMR) consistency test. The idea here is to estimate the final mass and spin of the remnant BH from the inspiral and merger-ringdown parts of the waveform independently assuming GR, and check the consistency between the two estimates. Although such a test was originally designed to perform a consistency test of GR, one can apply it to test specific modified theories of gravity as well. The third test is the GW propagation test, where we consider various non-GR effects related to GW propagation in a model-independent way, such as the modified dispersion relation of the graviton. Below, we will look at each of these tests in turn. We will use the geometric units of $c=G=1$ throughout.

\section{Parameterized Tests}\label{sec:parameterized}

One simple way of probing gravity with GWs in a theory-agnostic way is to introduce general, arbitrary deviations to the GR waveform template in the frequency domain, with the latter given by
\begin{equation}
   \tilde h_\GR(f)=A_\GR(f) e^{i\psi_\GR(f)}\,,
\end{equation}
where $A_\GR$ and $\psi_\GR$ are the GR amplitude and phase and $f$ is the GW frequency.
In such a framework, the GW community can simply probe arbitrary deviations from the predictions of GR with future GW observations without assuming an alternative theory a priori.
Once such constraints have been obtained, the results can finally be mapped to the large set of existing modified theories of gravity.

\subsection{Formalism}

\subsubsection{Parameterized Waveforms}

One method for performing model-independent tests of GR with GWs was proposed in~\cite{Arun:2006yw}. The inspiral portion of the waveform is given in terms of post-Newtonian (PN) expansion, where the relative velocity of the binary constituents are assumed to be much smaller than the speed of light. When each body is non-spinning, each PN term is given in terms of the masses $m_1$ and $m_2$. The authors proposed to measure each PN term independently and check the consistency between each other in the $m_1$-$m_2$ plane. This idea is similar to tests of GR with binary pulsar observations.
One downside of the above test is that one can only probe non-GR effect entering at PN orders where the GR terms are present, but there are many modified theories of gravity that predict the leading non-GR correction to enter at e.g. negative PN orders due to scalar radiation or the variation of the gravitational constant $G$ that are absent in the GR waveform. Another downside is that it may be difficult to perform similar tests with spinning compact binaries. 

\begin{figure}
\begin{center}
\includegraphics[width=0.8\columnwidth]{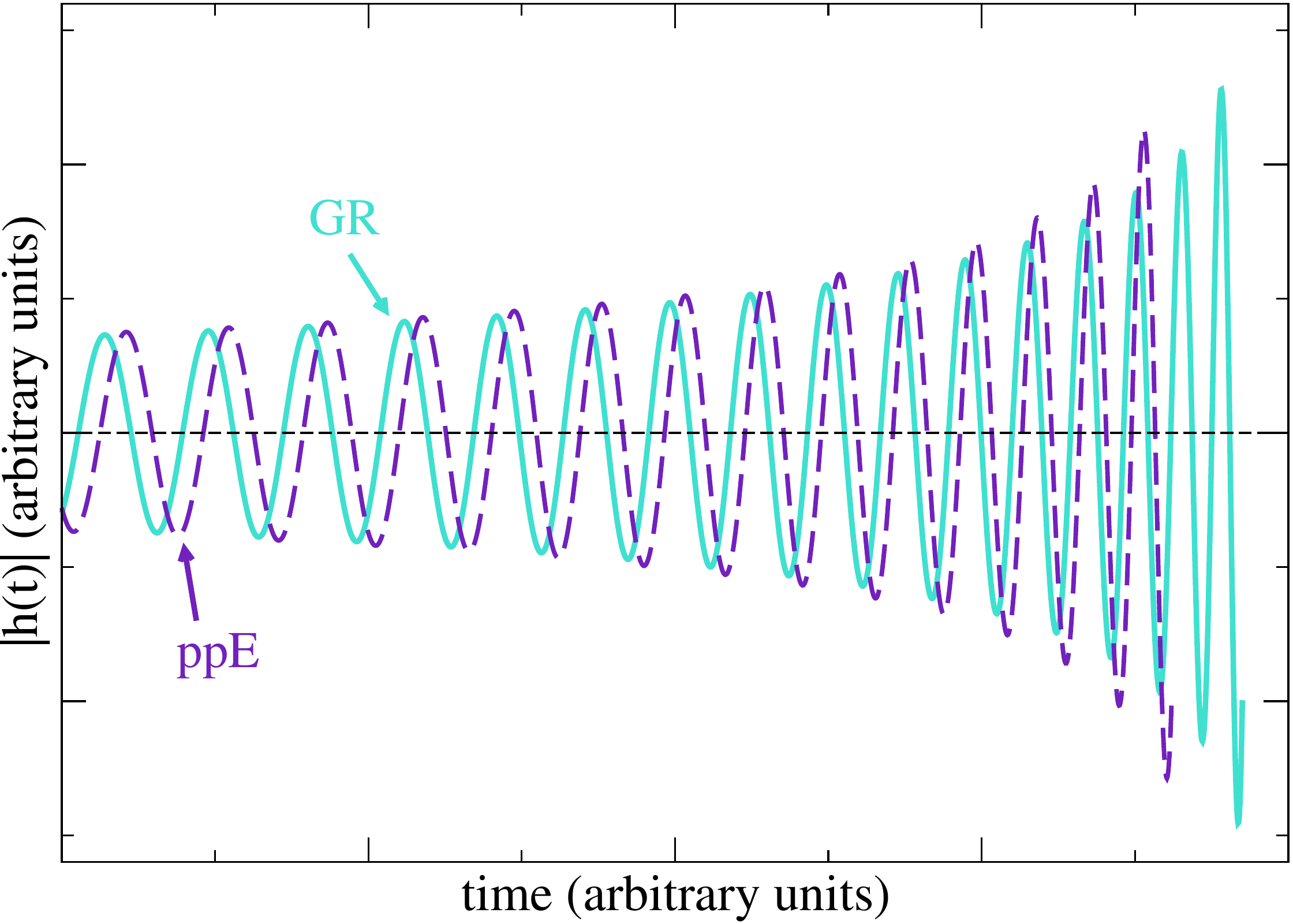}
\end{center}
\caption{Example waveforms of a compact binary inspiral in GR and ppE. Observe that they differ both in the amplitude and phase.
}\label{fig:ppE-waveform}
\end{figure}

To overcome these issues, the so-called \emph{parameterized post-Einsteinian} (ppE) formalism was proposed by Yunes \& Pretorius~\cite{Yunes:2009ke}, which allows one to characterize such arbitrary deviations to the GR amplitude and phase.
In particular, we take corrections of the form
\begin{equation}
\label{eq:PPE}
    A \to A_\GR(1+\alpha u^a), \hspace{7mm} \psi \to \psi_\GR+\beta u^b
\end{equation}
for effective relative velocity between our binary compact objects $u=(\pi \mathcal{M} f)^{1/3}$ with a chirp mass of $\mathcal{M}\equiv \frac{(m_1m_2)^{3/5}}{(m_1+m_2)^{1/5}}$.
From here on out, we shall refer to the new parameters $(\alpha,a)$ and $(\beta,b)$ as the amplitude and phase ppE parameters, respectively.
$\alpha$ and $\beta$ correspond to the \emph{size} of the modifications in question, and is what we will focus on constraining. 
$a$ and $b$ correspond to the power of velocity $u$ relative to GR at which a given correction alters the waveform.
These parameters define the PN order at which an effect alters the waveform like so
\begin{equation}
    a=2n, \hspace{7mm} b=2n-5,
\end{equation}
where an $n$-PN correction is proportional to $(u/c)^{2n}$ relative to the leading order GR term. 
The ppE modified waveform reduces to the GR one in the limit $(\alpha,\beta) \to (0,0)$. Figure~\ref{fig:ppE-waveform} shows an example comparison of the GR and ppE waveform.

LVC used a slightly different formalism called generalized inspiral-merger-ringdown (gIMR) formalism~\cite{Abbott_IMRcon2}. LVC first took the IMR phenomenological (IMRPhenom) GR waveforms that have various phenomenological parameters determined by fitting them with numerical relativity waveforms. LVC then allowed each of these parameters to vary from the GR one to account for generic non-GR effects. In the inspiral portion, there is a one-to-one mapping between PPE and gIMR formalisms~\cite{Yunes_ModifiedPhysics}. The latter also naturally includes corrections in the merger-ringdown portion. The original PPE formalism also has some proposals for the modified waveforms in the merger-ringdown portion~\cite{Yunes:2009ke}.

\subsubsection{Parameter Estimation}\label{sec:Fisher}

Now that we have built in arbitrary corrections to the GR waveform, we must work on constraining the ppE parameters with actual GW observations.
In practice, this is most commonly done by using a full statistical Bayesian analysis via e.g. Markov-chain Monte Carlo method to estimate posterior probability distributions on parameters, such as $\alpha$ and $\beta$ for various $a$ and $b$.
Working from the other end, one can derive the expressions for $(\alpha,a)$ and $(\beta,b)$ corresponding to various modified theories of gravity, which can then be used to map the constraints to theory-specific coupling parameters.
Such expressions can be found by first deriving corrections to the orbital separation and frequency evolution of a compact binary inspiral. These originate from corrections to the binary binding energy and GW luminosity due to e.g. modifications to the gravitational potential and radiation from additional channels like scalar and vector fields in the modified theory of gravity one wants to consider.
From here, one can calculate corrections to the GW amplitude and phase by following e.g.~\cite{Tahura_GdotMap}, where the resulting expressions for $(\alpha,a)$ and $(\beta,b)$ are displayed for a large list of modified theories of gravity that violate certain fundamental pillars of GR, including the equivalence principle, Lorentz/parity invariance and massless gravitons.

With this done, constraints on $\alpha$ and $\beta$ can be estimated across the entire PN spectrum for each present and future GW event, improving the constraints onward.
The resulting ruled-out regions of the parameter space can then be used to constrain the magnitude of various coupling parameters. 
As GW detection technology improves in the near future with upgraded ground-based and space-based interferometers, the space of allowable modified theories of gravity may shrink until the possibility of detecting beyond-GR behavior finally presents itself.

Another commonly-used method to predict posterior probability distributions on $\alpha$ and $\beta$ is known as the \emph{Fisher analysis} method~\cite{Cutler:Fisher}.
This analysis is a reliable approximation to the Bayesian analysis used by e.g. the LVC, for large signal-to-noise ratio (SNR) systems.
In particular the authors of Ref.~\cite{Yunes_ModifiedPhysics} found that for events with SNRs of $\sim 25$ (corresponding to that of GW150914), the Fisher analysis technique reliably reproduces the results of the Bayesian analysis done by the LVC for bounds on non-GR parameters.
Furthermore, as future GW interferometers become increasingly sensitive, GW events will become louder and louder, closing the gap between the Bayesian and Fisher analyses, with the latter being significantly less computationally expensive.

The Fisher analysis technique relies on the assumption that the inherent noise $n$ within the GW detector is distributed Gaussian, like so
\begin{equation}\label{eq:noise}
    p(n)\propto \exp\left\lbrack-\frac{1}{2}(n|n)\right\rbrack,
\end{equation}
where the inner product between two functions in the time domain $a(t)$ and $b(t)$ has been defined to be weighted by the detector's power spectral density $S_n(f)$ as
\begin{equation}
    (a|b)\equiv 2 \int\limits^{f_\mathrm{high}}_{f_\mathrm{low}}\frac{\tilde{a}^*\tilde{b}+\tilde{b}^*\tilde{a}}{S_n(f)}df.
\end{equation}
Here $\tilde a(f)$ and $\tilde b(f)$ are the corresponding function in the frequency domain, and * refers to complex conjugate. We have also assumed that the noise is stationary.
The upper and lower limiting frequencies $f_\mathrm{high}$ and $f_\mathrm{low}$ depend on the specific detector configuration used, and is described thoroughly in e.g.~\cite{Carson_multiBandPRD}.

The ultimate goal of the Fisher analysis is to find best fit parameters $\hat{\theta}^a$ (including ppE parameters $\alpha$, $\beta$) that whose template waveform $h$ maximally agrees with an observed signal $s$. 
By substituting in $s=n+h(\theta^i)$ into Eq.~(\ref{eq:noise}), we find the posterior probability distribution on $\theta^a$ to be
\begin{equation}\label{eq:maximalPDF}
    p(\theta^a|s)\propto p_{\theta^a}^{(0)}\exp\left\lbrack-\frac{1}{2}\Gamma_{ij}(\theta^i-\hat{\theta}^i)(\theta^j-\hat{\theta}^j)\right\rbrack.
\end{equation}
In the above expression, $p_{\theta^a}^{(0)}$ are the prior on parameters $\theta^i$, and $\Gamma_{ij}$ is the Fisher information matrix, defined to be
\begin{equation}
    \Gamma_{ij}\equiv \left(\frac{\partial h}{\partial \theta^i}\bigg| \frac{\partial h}{\partial \theta^j}\right).
\end{equation}
Equation~(\ref{eq:maximalPDF}) corresponds to a multi-variate Gaussian probability distribution and the resulting root-mean-square errors on parameters $\theta^a$ are given simply as
\begin{equation}
    \Delta\theta_i=\sqrt{\Gamma_{ii}^{-1}}.
\end{equation}
The prior distribution is arbitrary, though in practice, the outcome results is kept simple if one uses a Gaussian prior. For example, if one desires to include Gaussian prior distributions with a standard deviation $\sigma_{\theta^i}^{(0)}$ on template parameters or to combine the results of multiple observations on detectors $A=1 \dots N$, the Fisher information matrix simply becomes 
\begin{equation}
    \Gamma_{ij}\to\tilde\Gamma_{ij}=\frac{1}{\left(\sigma_{\theta^i}^{(0)}\right)^2}\delta_{ij}+ \sum_{A=1}^N\Gamma_{ij}^A\,,
\end{equation}
where $\Gamma_{ij}^A$ is the Fisher matrix of the $A$th detector.

\subsection{Current Status}
Now that we have built up the formalism of parameterized tests of GR, let us discuss their present considerations from current GW observations.
In this section we will mainly focus on constraints formed from the first two binary BH GW observations of GW150914 and GW151226~\cite{GW_Catalogue}. 
The former event was located a distance of $430^{+150}_{-170}$ Mpc away with a large SNR of 25.1, with constituent masses of $35.6^{+4.8}_{-3.0}\mathrm{ M}_\odot$ and $30.6^{+3.0}_{-4.4}\mathrm{ M}_\odot$ and dimensionless effective spin of $-0.01^{+0.12}_{-0.13}$
(here, the dimensionless effective spin parameter is the mass-weighted average of compact objects' angular momentum component normal to the orbital plane divided by their masses squared).
The latter event was located $440^{+180}_{-190}$ Mpc away, was detected with a smaller SNR of 13, and was comprised of $13.7^{+8.8}_{-3.2}\mathrm{ M}_\odot$ and $7.7^{+2.2}_{-2.6}\mathrm{ M}_\odot$  BHs with an effective dimensionless spin of $0.18^{+0.2}_{-0.12}$.

For the remainder of this section, we focus only on constraints on the ppE phase parameter $\beta$.
As found by Tahura \emph{et al} in~\cite{Tahura:2019dgr}, the inclusion of the ppE amplitude parameter only alters constraints on $\beta$ by $\sim 10\%$.
Thus, we and many other authors consider constraints on the phase correction only.

\begin{figure}
\begin{center}
\includegraphics[width=0.8\columnwidth]{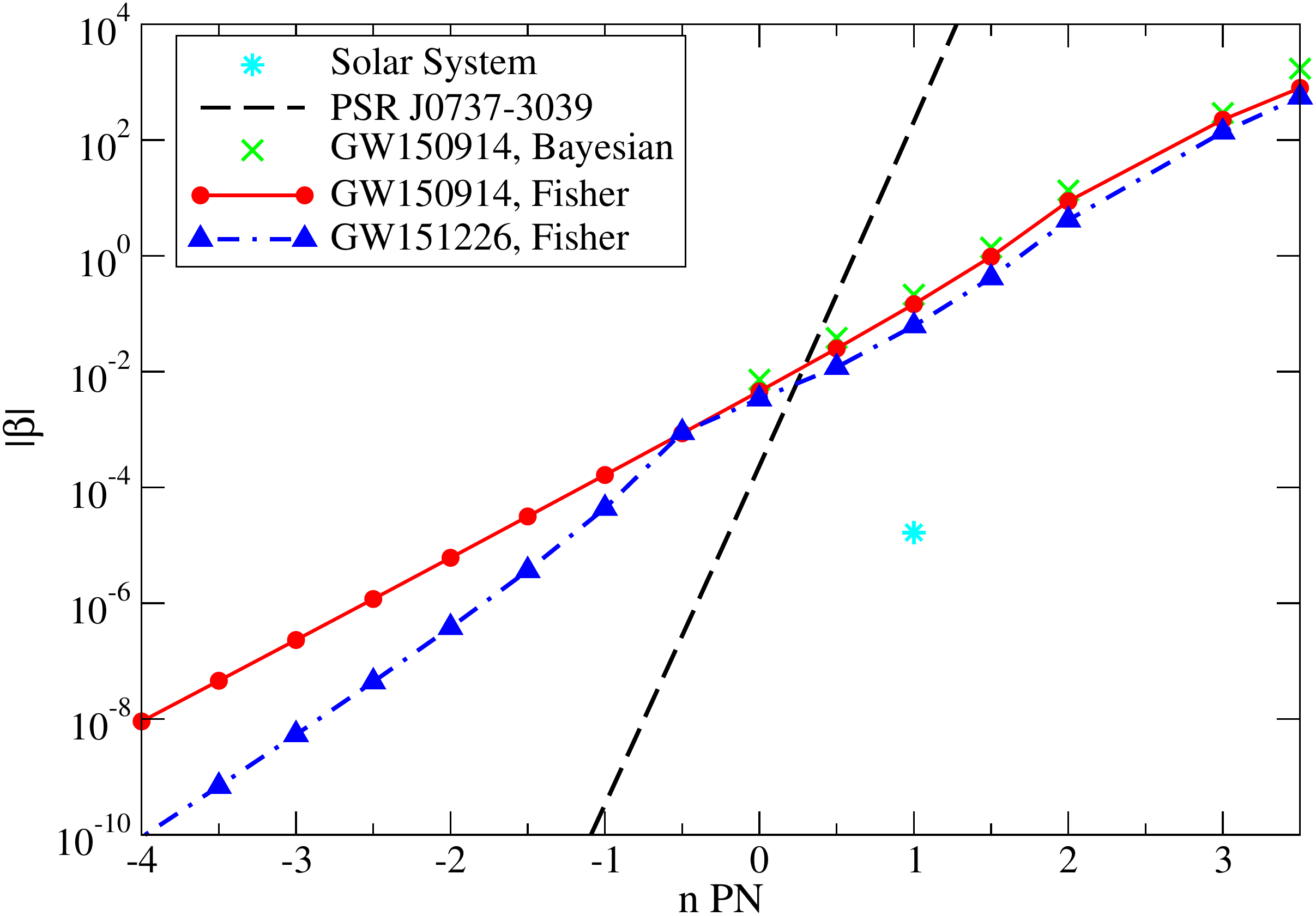}
\end{center}
\caption{The 90\% credible upper bounds on the ppE parameter $\beta$ (Eq.~\eqref{eq:PPE}) at each PN order the correction enters. We show bounds using {solar system experiments (cyan star), binary pulsar observations~\cite{Yunes:2010qb} (black dashed),} GW150914 with Bayesian~\cite{Abbott_IMRcon} (green crosses) and Fisher~\cite{Yunes_ModifiedPhysics} (red solid) analyses, and GW151226 with a Fisher analysis~\cite{Yunes_ModifiedPhysics} (blue dotted-dashed). 
This figure is taken from~\cite{Yunes_ModifiedPhysics}.
}\label{fig:ppE}
\end{figure}

We first turn our attention to the work done by the LVC in~\cite{Abbott_IMRcon} for a catalog of binary BHs in~\cite{GW_Catalogue} and for the binary neutron star (NS) merger event GW170817 in~\cite{Abbott:2018lct}.
In these works, the collaboration 
considered several tests of GR using a Bayesian analysis framework.
For all cataloged events thus far, none of them deviate from GR to a statistically significant degree.
This does not indicate that no deviations exist - they may simply be hidden in the relatively large detector noise which will only improve as time continues. 
Figure~4 of~\cite{Abbott_IMRcon} and Fig.~2 of~\cite{Abbott:2018lct} present constraints on the arbitrary phase deviation parameter for each event considered across the spectrum of PN orders. The ones for GW150914 are shown as green crosses in Fig.~\ref{fig:ppE}.

Figure~\ref{fig:ppE} also displays the  90\% upper-limit constraints on $\beta$ as a function of PN order from both GW150914 and GW151226~\cite{Yunes_ModifiedPhysics} via a Fisher analysis, showing not only bounds on the positive PN orders but also on negative PN ones.
Additionally shown are the results from the previous strongest solar-system constraints and constraints from the double pulsar system PSR J0737-3039~\cite{Yunes:2010qb}.
We observe several interesting details from the listed figure: 
\begin{enumerate}
    \item The Bayesian and Fisher analysis on GW150914 agree with each other very well, confirming that the latter is valid as an order of magnitude estimate.
    \item GW150914 and GW151226 produce similar results for positive-PN orders, while GW151226 places stronger bounds than GW150914 for negative-PN orders by up to two orders of magnitude. 
    \item Solar system bounds outperform GW observations for 1PN order only. Though a care must be taken in this comparison since the former only includes conservative effects (such as corrections to the Keplerian motion) while the latter includes both conservative and dissipative (GW emission) effects.
    \item The binary pulsar observation from PSR J0737-3039 produces significantly stronger results at negative PN orders than the GW observations, and vice versa for positive PN orders. This binary pulsar bound only takes into account the dissipative effects.
\end{enumerate}
However, we note that while final two observations from the above list produce stronger constraints at certain PN orders than GW observations, they originate from weak and/or static field environments, thus the GW constraints remain unique. 
Moreover, certain modified theories of gravity induce large corrections only to BHs, and hence it is important to derive bounds from various sources.

\subsection{Future Prospects}

\begin{figure}[htb]
\begin{center}
\includegraphics[width=.7\linewidth]{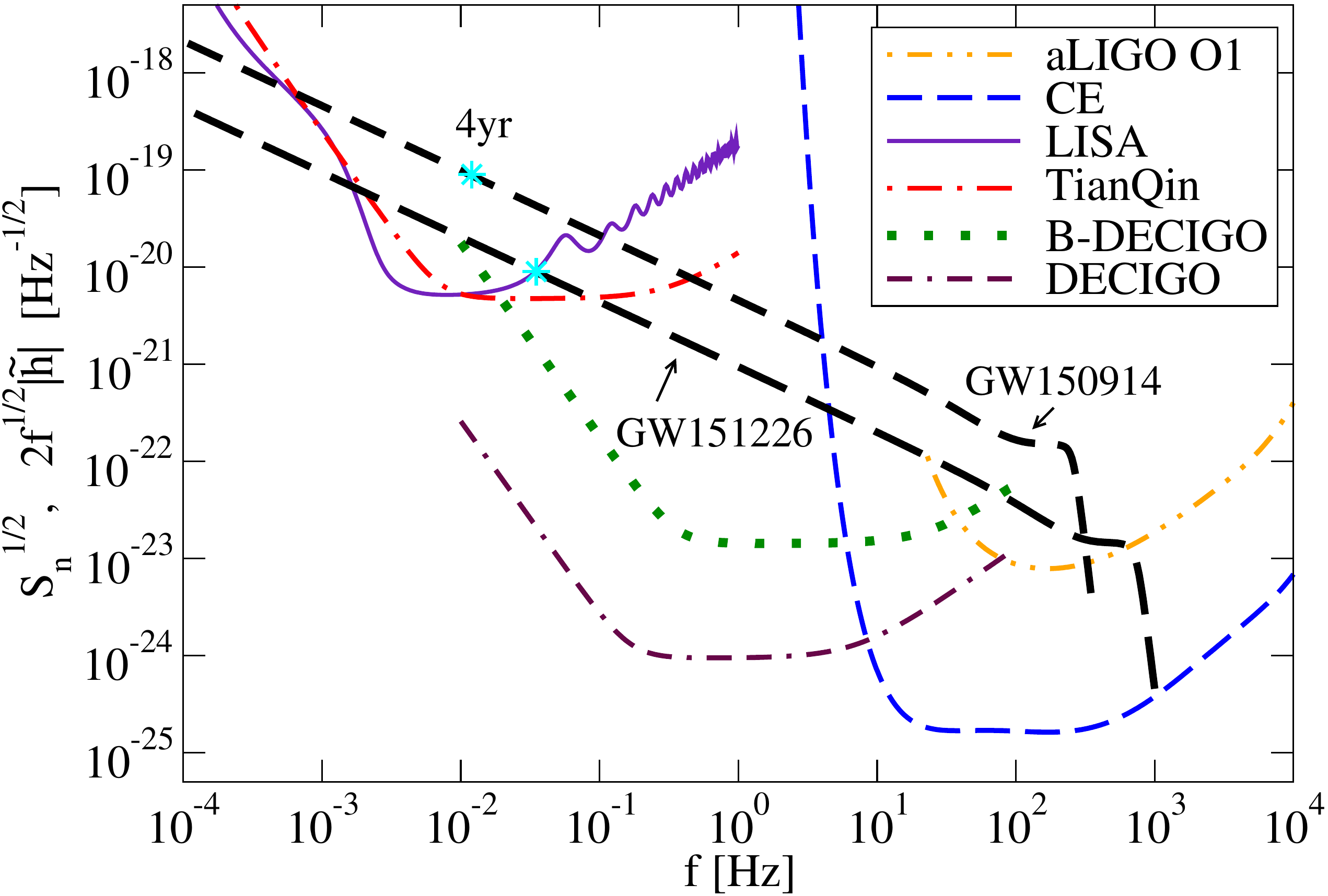}
\end{center}
\caption{
(Square root of) spectral noise densities $\sqrt{S_n(f)}$ of various gravitational-wave interferometers.
We also show the characteristic amplitudes $2\sqrt{f}|\tilde h(f)|$ for GW150914 and GW151226, together with 4 years prior to merger shown as cyan stars. The ratio between the GW spectrum and noise sensitivity roughly corresponds to the SNR of the event.
The early inspiral portions of the BH coalescences are observed by the space-based detectors, while the late inspiral and merger-ringdown portions are observed by the ground-based detectors. This figure is taken from~\cite{Zack:Proceedings}.
}\label{fig:detectors}
\end{figure}

Now let us consider predictions we can make on the future of parameterized tests of GR.
Typically, this is accomplished by choosing future-generation GW detector sensitivity predictions (see Fig.~\ref{fig:detectors}), and using a Fisher or a Bayesian analysis to predict constraints on the theory-agnostic ppE phase and/or amplitude parameters $\alpha$ and $\beta$.

\begin{figure}[htb]
\begin{center}
\includegraphics[width=.7\linewidth]{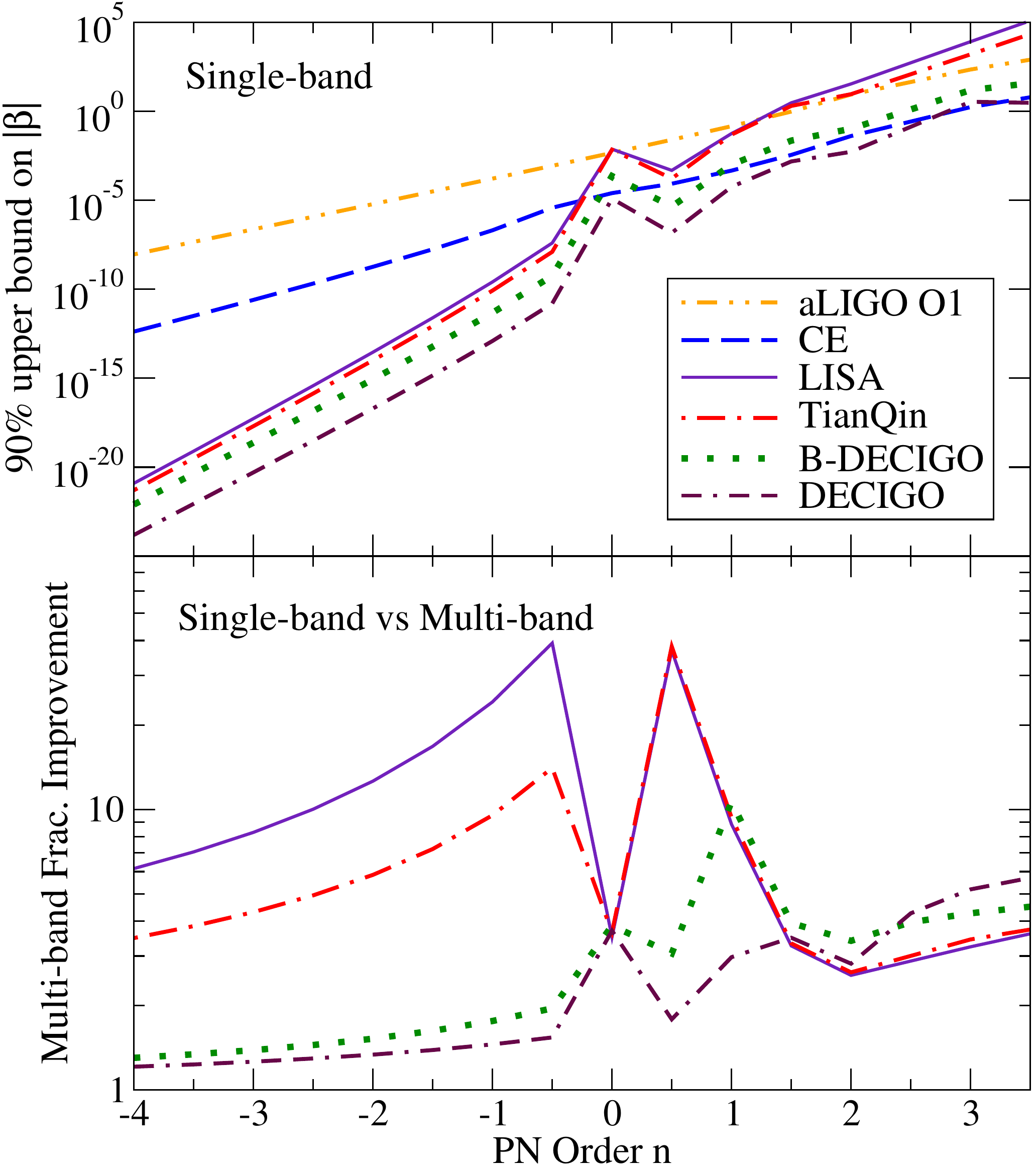}
\end{center}
\caption{(Top) 90\% credible upper bounds on $\beta$ as a function of the PN order that a non-GR correction enters assuming that each interferometer detects GW signals from a GW150914-like event.
(Bottom) Improvement on the bound on $\beta$ with multi-band GW observations (each space-based detector + CE) compared to the single band case. 
This figure is taken and edited from~\cite{Carson_multiBandPRL}. 
}\label{fig:future_beta}
\end{figure}

The top panel of Fig.~\ref{fig:future_beta} summarizes how the bound on $\beta$ improves with future GW interferometers, assuming that they detect GW signals from a GW150914-like event. Observe that LISA has a significant improvement, especially when corrections enter at negative PN orders (similar for TianQin). This is because the relative velocity of BHs is much smaller for LISA than ground-based detectors, and thus (relative) negative-PN effects become enhanced (if they exist). Observe also that DECIGO outperforms all the interferometers due to its high detector sensitivity and a large number of GW cycles during its observation.

One particularly promising line of inquiry is into the potential \emph{multi-band} observations~\cite{Sesana:2016ljz} of GWs from GW150914-like stellar-mass BH binaries between both space-based detectors (such as LISA, TianQin, (B-)DECIGO, and many others) and ground-based detectors (such as the Cosmic Explorer (CE), Einstein Telescope (ET), and many others). Figure~\ref{fig:detectors} shows the GW spectra for GW150914, which can be detected both ground-based and space-based interferometers.
Such an observation would take place during both the low-frequency bands ($\sim 10^{-4} \text{ Hz}-1 \text{ Hz}$) when the (stellar-mass) BHs are far apart and moving slowly, as well as the high-frequency bands ($\sim 1 \text{ Hz}-10^{4} \text{ Hz}$) when the BHs are moving rapidly, making contact, and ringing down to their final state.
By combining together the entire observation, we can expect stronger bounds to be placed on $\alpha$ and $\beta$ along all PN orders, and corresponding constraints on most modified theories of gravity. 
For example, the prospects of probing scalar dipole radiation (absent in GR) with multi-band observations are discussed in~\cite{Barausse:2016eii}.

The bottom panel of Fig.~\ref{fig:future_beta} shows the improvement on the bound on $\beta$ using multi-band observations with respect to the single band case. Observe that LISA+CE may have an improvement over LISA-alone or CE-alone by up to a factor of 40. On the other hand, DECIGO does not have much improvement with the multi-band observations because its sensitivity is already good enough on its own.
This predictive work is extremely valuable as they can strengthen the case for the development of future detectors, and can prove to be informative to decisions based on detector design.

\subsection{Applications to Specific Theories}

The above bounds on the arbitrary ppE phase parameter can further be mapped to desired modified theories of gravity by considering the mappings displayed in e.g.~\cite{Tahura_GdotMap}. The results are summarized in Table~\ref{tab:summary} for the following theories:
\begin{itemize}
\item \emph{Einstein-dilaton Gauss-Bonnet (EdGB) gravity}: A scalar field (dilaton) is coupled to a Gauss-Bonnet combination of the Riemann curvature squared in the action with a coupling constant $\alpha_\EdGB$, motivated by string theory.
\item \emph{scalar-tensor theories}: Generic theories with a scalar field coupled either minimally or non-minimally to the metric. BHs can acquire a scalar charge when the scalar field is time evolving with a rate $\dot \phi$.
\item \emph{dynamical Chern-Simons (dCS) gravity}: A (pseudo) scalar field is coupled to the Pontryagin density (odd-parity curvature squared scalar) in the action with a coupling constant $\alpha_\dCS$, motivated by string theory, loop quantum gravity and effective field theory of inflation. 
\item \emph{noncommutative gravity}: Quantizing spacetime by promoting the coordinates $x^\mu$ to operators $\hat x^\mu$. For example, $[\hat t, \hat x^i] = i \theta^{0i} $ where $\theta^{0i} \theta^{0}{}_i = \Lambda^2 l_p^2 t_p^2$ with $\Lambda$ being the noncommutative parameter and $l_p$ and $t_p$ representing the Planck length and time respectively.
\item  \emph{time-varying mass theories}: Theories in which BH masses may change over time with a rate $\dot M$ due to e.g. enhanced Hawking radiation in some extra dimension models. Time-variation for BH masses are also relevant in astrophysical/cosmological context in terms of accretion of gas and dark energy.
\item \emph{time-varying $G$ theories}: Theories in which $G$ varies over time with a rate $\dot G$, such as scalar-tensor theories.
\item \emph{massive graviton theories}: Theories in which the graviton has a non-vanishing mass $m_g$. The graviton mass affects the GWs both during generation and propagation.
\end{itemize}
Notice that GW bounds are typically much weaker than other bounds, though the former have a meaning that they are the bounds obtained in the strong and dynamical field regime. Having said this, there are some theories, like noncommutative gravity, on which GW events placed stringent bounds.

{
\newcommand{\minitab}[2][l]{\begin{tabular}{#1}#2\end{tabular}}
\renewcommand{\arraystretch}{1.4}
\begingroup
\begin{table*}[t]
\begin{centering}
\resizebox{\textwidth}{!}{%
\begin{tabular}{ccccccc}
\hline
\hline
\noalign{\smallskip}
 {\bf{Theory}} & {\bf{GR Pillar}} & {\bf{PN}} & {\bf{Repr. Par.}} & {\bf{GW150914}}& {\bf{GW151226}}  & {\bf{Other Bounds}} \\
\hline \hline
 EdGB & \multirow{2}{*}{SEP}  & \multirow{2}{*}{$-1$}  & $\sqrt{|\alpha_\EDGB|}$ [km] & 
 --- & 5.6 &  $10^7$, 2\\
 scalar-tensor  &   & & $|\dot{\phi}|$ [1/sec] & 
 --- & $1.1 \times 10^4$ &    $10^{-6}$ \\ 
 \hline
dCS & SEP & \multirow{2}{*}{$+2$} & $\sqrt{|\alpha_\CS|}$ [km]  & --- &--- &   $10^8$\\ 
 noncommutative  & commutativity &   &  $\sqrt{\Lambda}$ & 3.5 &--- &   ---\\ 
\hline 
time-varying $M$  & 4D & $-4$ & $\dot M$ [$M_\odot$/yr] & $4.2\times10^8$& $5.3\times10^6$ & ---\\
\hline
\multirow{2}{*}{time-varying $G$}  & \multirow{2}{*}{SEP}& \multirow{2}{*}{$-4$} &  \multirow{2}{*}{$|\dot G|$ [10$^{-12}$/yr]} & $5.4 \times 10^{18}$ & $1.7 \times 10^{17}$ & \multirow{2}{*}{0.1--1}\\ 
& & & & $7.2 \times 10^{18}$ & $2.2 \times 10^{16}$ & \\
\hline
\multirow{2}{*}{massive graviton}  & \multirow{2}{*}{$m_g=0$} & \multirow{1}{*}{$-3$}  & \multirow{2}{*}{$m_g$ [eV]} & \multirow{1}{*}{$6.4\times10^{-14}$} & \multirow{1}{*}{$10^{-14}$, $3.1\times10^{-14}$} &  \multirow{1}{*}{$10^{-21}$--$10^{-19}$} \\
 & & \multirow{1}{*}{$+1$}  &  &   \multirow{1}{*}{$10^{-22}$} & \multirow{1}{*}{$2.9 \times 10^{-22}$} &  \multirow{1}{*}{$10^{-30}$--$10^{-23}$}   \\
\noalign{\smallskip}
\hline
\hline
\end{tabular}
}
\end{centering}
\caption{Summary of constraints on example modified theories of gravity with GW150914 and GW151226, together with other bounds. (1st column) example theories; (2nd column) violation of GR fundamental pillars, such as the Strong Equivalence Principle (SEP), four-dimensional spacetime (4D), and massless gravitons ($m_g=0$); (3rd column) the PN order at which the leading correction enters; (4th column) representative theoretical parameters; (5th column) bounds on these representative parameters with GW150914; (6th column) same as the 5th column but for GW151226; (7th column) bounds from other observations.
The top (bottom) row within massive graviton corresponds to corrections to the dynamical (propagation/conservative) sector.
This table is taken and edited from~\cite{Yunes_ModifiedPhysics,Zack:Proceedings}.
 }
\label{tab:summary}
\end{table*}
\endgroup
}

We now move onto describing future prospects on constraining example theories via parameterized tests.
We first take a look at the work done by Chamberlain \& Yunes in~\cite{Chamberlain:2017fjl}.
Here, the authors estimated constraints on several modified theories of gravity (including dipole radiation, extra dimensions, time-varying gravitational constant, Einstein-\AE ther, khronometric, and massive graviton theories of gravity) obtained with a Fisher analysis for a variety of astrophysical systems and ground- and space-based detectors.
In particular, they also considered several alternative design sensitivities for the space-based detector LISA to compare the resulting constraints.
Table~I of the same paper neatly displays the best constraints one can place on each theory of gravity considered from ground-based or space-based interferometers, including the current constraints, as well as the best astrophysical systems for placing constraints with both ground-based and space-based detectors.

\begin{figure}
\begin{center}
\includegraphics[width=.48\columnwidth]{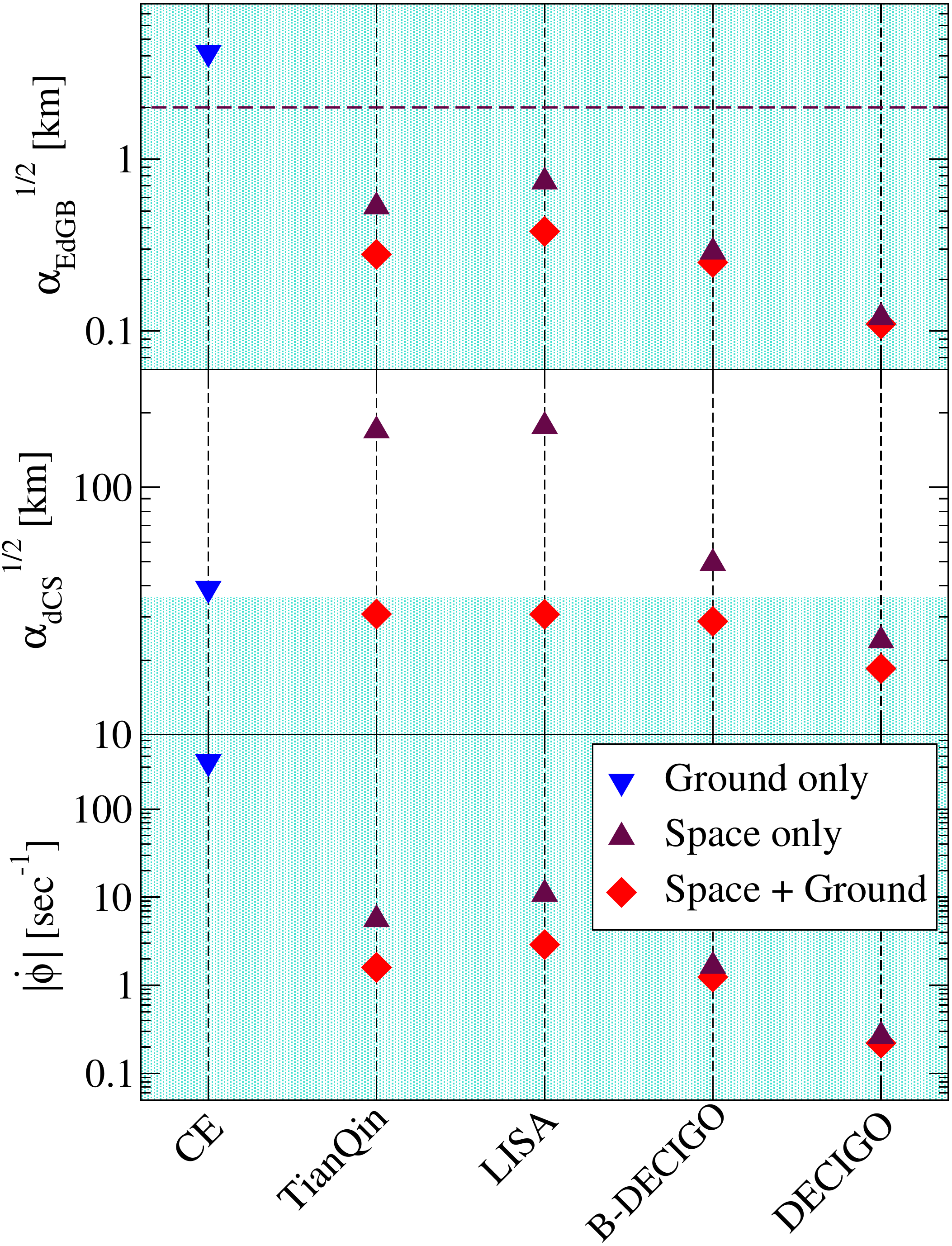}
\includegraphics[width=.48\columnwidth]{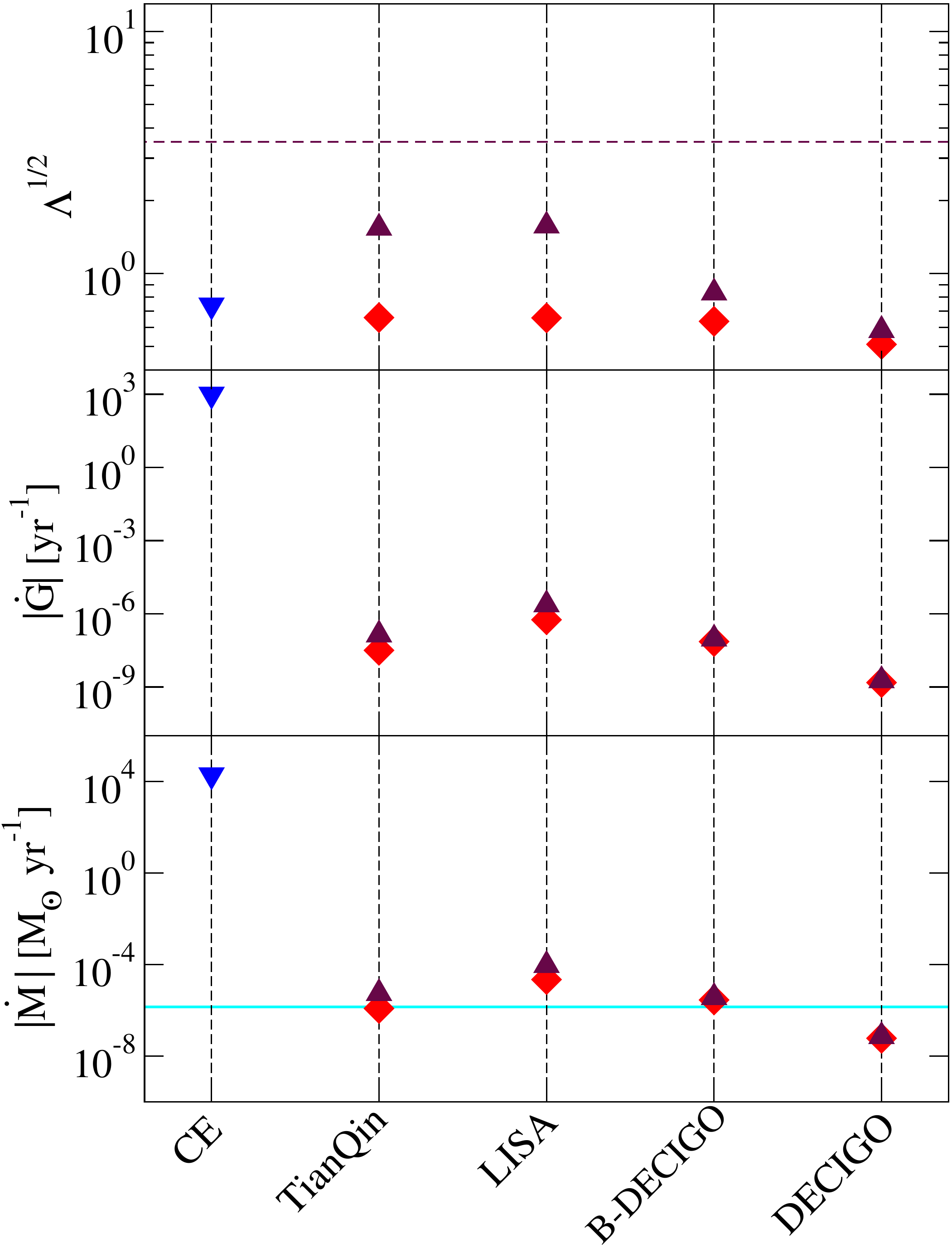}
\end{center}
\caption{Estimated projections on the 90\% credible-level upper bounds on the theoretical parameters representative of the theories considered in Table~\ref{tab:summary} (except for the graviton mass that will be discussed in Sec.~\ref{sec:mg}) with GW150914-like events using future GW detectors. We compare bounds with ground-based detector (CE) alone, space-based ones alone and multi-band observations.
The blue shaded regions correspond to the area where the small-coupling approximation (that is necessary to treat the theory as effective theory) is valid, and the dashed maroon lines correspond to the current constraints found in the literature.
The horizontal cyan line in the last panel corresponds to the Eddington accretion rate for BHs in GW150914-like events that corresponds to the maximum accretion rate under spherical symmetry. This figure is taken from~\cite{Carson_multiBandPRD}. 
}\label{fig:Muiltiband}
\end{figure}

Figure~\ref{fig:Muiltiband} presents the future predicted bounds on various theoretical parameters in theories mentioned earlier with GW150914-like events with CE and space-based detectors (LISA, TianQin, B-DECIGO, and DECIGO) computed by ourselves~\cite{Carson_multiBandPRD}.
Observe that space-based detectors place stronger bounds than CE for theories in which the leading correction enters at negative PN orders (EdGB, scalar-tensor, varying-$G$ and varying-mass theories). This is because such corrections (relative to GR) become larger at lower frequencies when the relative velocity is smaller. On the other hand, CE places stronger bounds than the space-based detectors for theories with positive-PN corrections (dCS and noncommutative theories).
Here we can also see several cases where ground-based and/or space-based observations can provide constraints stronger than the current ones found in the literature.
Figure~\ref{fig:Muiltiband} also presents the results for multi-band observations.
We  see that in every scenario, the multi-band observations produce stronger bounds than both the space-based or ground-based ones individually.
Finally, we observe that in dCS gravity, both of the ground-based and space-based detectors (except for DECIGO) fail to provide constraints consistent with the theories' small-coupling approximation (needed to treat the theory as a valid effective theory), and are thus not reliable.
However, only when the multi-band observations are made is this approximation satisfied and new constraints can be placed that are stronger than the current ones by seven orders of magnitude.

We end this section by mentioning future prospects for testing GR with mixed NS/BH binaries. Several candidates for such mixed binaries were reported during the O3 run. These binaries are especially useful for probing theories with scalar fields. This is because such theories generically allow the presence of the scalar dipole radiation, which is proportional to the square of the difference in the scalar charges of the binary constituents. Thus, the amount of scalar radiation becomes larger for binaries consisting of different types of compact objects.

\begin{figure}[h]
\begin{center}
\includegraphics[width=0.8\columnwidth]{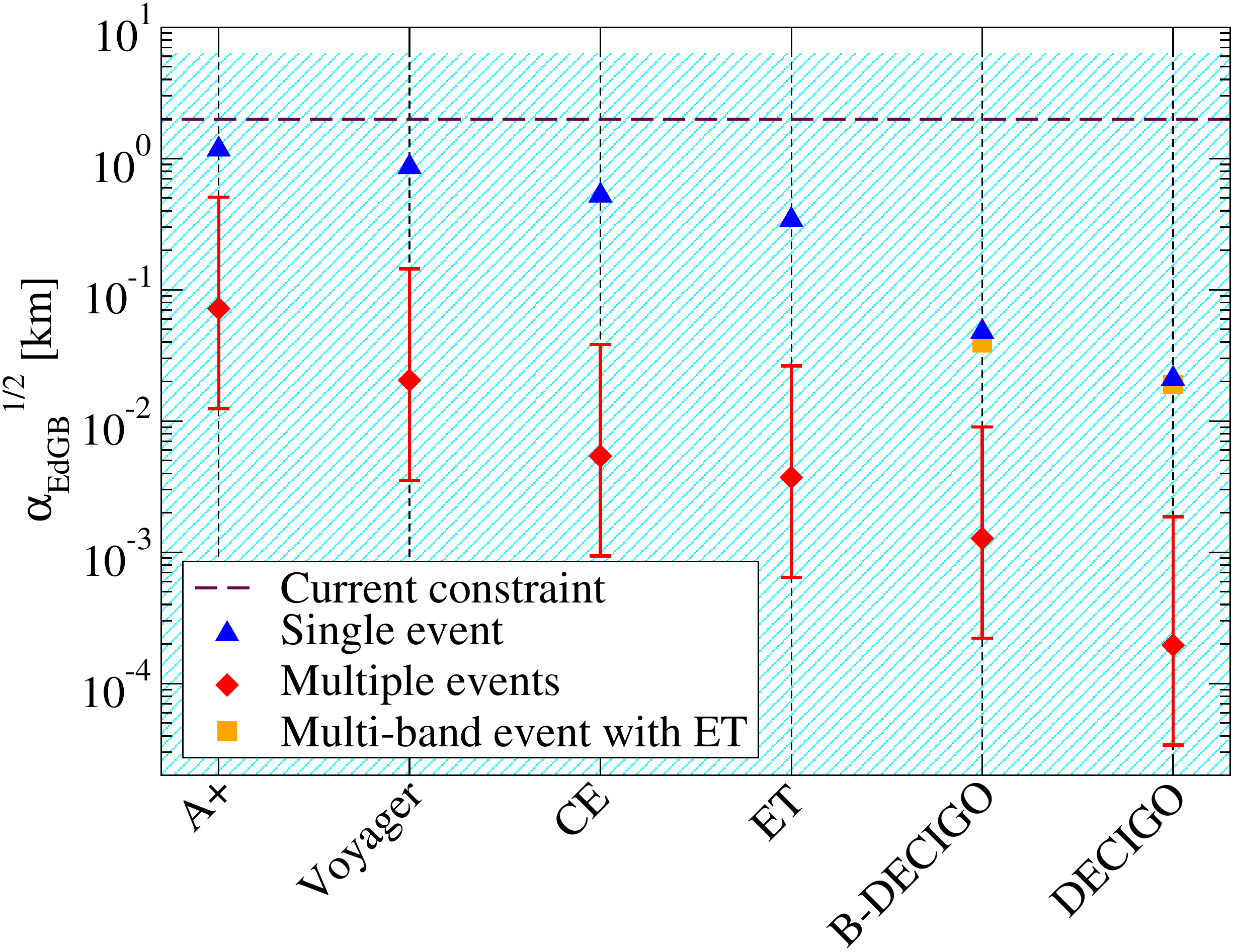}
\end{center}
\caption{
Future projected bounds on the EdGB parameter $\sqrt{\alpha_\EdGB}$ with NS/BH binaries using various future GW detectors. We show bounds with both single and multiple events. The error bars on the latter are due to uncertainty in the number of NS/BH detections with future observations. For space-based detectors, we also show bounds with multi-band observations. The current constraint is shown by the dashed horizontal line while the blue shaded region corresponds to the region where the small coupling approximation is valid. Observe that the future observations may be able to strengthen the bound by up to four orders of magnitude relative to the current bound. This figure is taken from~\cite{Zack:mixedBinaries}.
}\label{fig:NSBH}
\end{figure}

To put the discussion into context, let us here focus on EdGB gravity as an example, which we can probe within the ppE framework by looking at the -1PN correction bound. In~\cite{Zack:mixedBinaries}, we first studied the prospects of probing this theory with the O3 NS/BH candidates. We found that if the BH mass is smaller than $\sim 16.5M_\odot$, it is likely that one can place a bound that is stronger than the current bound. Figure~\ref{fig:NSBH} presents the projected future bounds on the EdGB coupling constant with NS/BH binaries using various ground-based and space-based GW detectors (including multi-band observations). We show the cases for both single events and multiple events. Observe that the bound can become stronger than the current one by up to four orders of magnitude. This, in turn, means that if the true theory of gravity is EdGB and the true coupling constant lies somewhere between $10^{-4}\mathrm{km}<\sqrt{\alpha_\EdGB} < 2\mathrm{km}$, future NS/BH GW observations have the potential to detect the EdGB correction encoded in the waveform.

\section{Inspiral-merger-ringdown Consistency Tests}

Now let us consider another model-independent test of GR, this time by comparing the consistency between the inspiral and merger-ringdown portions of a given GW signal.
This test is most commonly known as the IMR consistency test and has been widely used in the community to generically test arbitrary GW signals for any signs of deviation from GR.
Somewhat unlike the parameterized tests considered in the previous section, the IMR consistency tests allow us to to detect any arbitrary deviations residing within the entire GW signal, while the former allows one to constrain deviations at pre-specified PN orders in the inspiral portion\footnote{LVC used the gIMRPhenom formalism to probe parameterized deviations in the merger-ringdown as well.}.
In the IMR consistency tests, we will begin by checking the consistency between the inspiral GW signal when the compact objects are largely separated and moving relatively slow, and the merger-ringdown signal when the objects first make contact and form a remnant BH which settles down to its final state via the radiation of quasi-normal modes (QNMs).
The separation between the two signals in question is commonly defined to be when the compact objects reach the location of the inner-most stable circular orbit (ISCO) (e.g. when the GW frequency reaches $f_\ISCO$), at which point they enter a plunging orbit and finally make contact with each other.
If the two signals agree with each other to a statistically significant level within the detector noise using GR template waveforms, we will say that GR is consistent and no deviations from it can be found.
On the other hand, if deviations between the two signals are found, one can claim a general deviation from GR present somewhere within the observed signal.
Further follow-up studies can then be performed to determine the precise nature of the deviation, and if it can be mapped to any known alternative theories of gravity.

The consistency between the inspiral and merger-ringdown signals is performed by simply estimating the remnant BH's final mass  $M_f$ and (dimensionless) spin $\chi_f$ given the observation of only the former or latter signal, without the presence of the other. 
This is estimated by using the fit for $M_f$ and $\chi_f$ in terms of the initial masses and spins obtained from numerical relativity simulations, as we will discuss in more detail later.
We can then draw the two-dimensional posterior probability distribution functions between the remnant BH's mass and spin from both of the inspiral and merger-ringdown portions of the signal.
The contours are then compared with each other
to determine the consistency between the two portions of the GW signal.
This is commonly done by transforming to the new coordinates $\propto(\Delta M_f, \Delta \chi_f)$, where $\Delta M_f$ and $\Delta \chi_f$ describe the differences between the inspiral and merger-ringdown predictions of the final mass and spin. 
Thus, finally the compatibility of the resulting contour with the ``perfect-match'' value of $(0,0)$ gives us a reliable test of GR which allows us to detect even minor deviations from GR, depending on the sensitivity of the GW detector in use.

\subsection{Formalism}\label{sec:IMRformalism}
Let us now discuss the various technicalities of the IMR consistency test~\cite{Ghosh_2017} as is commonly used by the LVC and others.
As discussed above, given an observed (or simulated) GW signal with sufficient SNR in \emph{both} of the inspiral (I) and merger-ringdown (MR) portions\footnote{This corresponds to an SNR of at least 9 in both portions, as was achieved in the first GW observation GW150914~\cite{GW150914}, which was measured with SNRs of 19.5 and 16, respectively.}, which are separated by the ISCO frequency $f_{\ISCO}=(6^{3/2} \pi M)^{-1}$ for total binary mass $M\equiv m_1+m_2$, we separate the signal into two distinct regions: ``I'' ($f<f_\ISCO$), and ``MR'' ($f>f_\ISCO$).
If GR is correct, the remnant BH's mass and spin can be uniquely predicted entirely from the intrinsic masses and spins of the initial BHs by the no-hair theorem, namely $M_f=M_f(m_1,m_2,\chi_1,\chi_2)$ and $\chi_f=\chi_f(m_1,m_2,\chi_1,\chi_2)$.
Such expressions are quite complicated and have been determined through various numerical relativity simulations of binary BH mergers, e.g. in~\cite{PhenomDII} where the authors provided fits for $M_f$ and $\chi_f$ within the IMRPhenomD gravitational waveform from a wide variety of binary BH simulations.

Using such expressions, we can begin our process by first estimating posterior probability distributions on the initial parameters $m_1^{\I,\MR}$, $m_2^{\I,\MR}$, $\chi_1^{\I,\MR}$, and $\chi_2^{\I,\MR}$ from the inspiral and merger-ringdown portions of the signal, individually.
Such distributions are most reliably reconstructed using a Bayesian analysis as is done by the LVC. However, as we discuss later, this can also be approximated by the Fisher analysis techniques described in Sec.~\nameref{sec:Fisher}.
While the latter method is an approximation to the former, it becomes especially useful for the forecasting of future IMR consistency tests of GR from signals which have yet to be detected, for which the SNRs are expected to be high and the validity of the approximation increases.
With the above probability distributions on the initial masses and spins in hand, we can then transform them into posterior probability distributions on the remnant BH's mass and spin $M_f^{\I,\MR}$ and $\chi_f^{\I,\MR}$ from the inspiral and merger-ringdown signals using e.g. the numerical relativity fits of~\cite{PhenomDII}, or from any other chosen gravitational waveform model.

Typically, to more qualitatively judge the comparison between the inspiral and merger-ringdown, we transform the two contours into a new shared coordinate system.
Namely, we choose the new coordinates:
\begin{equation}\label{eq:coordinates}
\epsilon\equiv\frac{\Delta M_f}{\overline{M}_f}, \hspace{5mm} \sigma\equiv\frac{\Delta \chi_f}{\overline{\chi}_f}.
\end{equation}
Here $\Delta M_f \equiv M_f^\I-M_f^\MR$ and $\Delta \chi_f \equiv \chi_f^\I-\chi_f^\MR$ describe the differences between the inspiral and merger-ringdown mass and spin predictions, and $\overline{M}_f \equiv \frac{1}{2}(M_f^\I+M_f^\MR)$, $\overline{\chi}_f \equiv \frac{1}{2}(\chi_f^\I+\chi_f^\MR)$ describe their averages.
In this new coordinate system, we are allowed a simple way to directly compare the two probability distributions with only a single contour.
In fact, notice how the condition for a perfect match between the two obeys the point $(\epsilon,\sigma)=(0,0)$, corresponding the GR being the true theory of gravity found in nature.
To transform the inspiral ($P_\I$) and merger-ringdown ($P_\MR$) probability distributions in the $(M_f,\chi_f)$ plane to the new $(\epsilon,\sigma)$ coordinate system, we simply apply the following transformation found in the Appendix of Ref.~\cite{Ghosh_2017}:
\begin{align}\label{eq:transform}
\nonumber P(\epsilon,\sigma)=&\int\limits^1_0 d\bar{\chi}_f \int\limits^{\infty}_0 d\bar{M}_f P_\I \left( \left\lbrack 1+\frac{\epsilon}{2} \right\rbrack \bar{M}_f , \left\lbrack 1+\frac{\sigma}{2} \right\rbrack \bar{\chi}_f \right) \\
&\times P_\MR  \left( \left\lbrack 1-\frac{\epsilon}{2} \right\rbrack \bar{M}_f , \left\lbrack 1-\frac{\sigma}{2} \right\rbrack \bar{\chi}_f \right) \bar{M}_f \bar{\chi}_f  .
\end{align}
The compatibility of the new probability distribution with the value of $(0,0)$ determines the success or failure of the IMR consistency test, at various statistical confidence intervals.

Finally, let us consider how one would utilize the IMR consistency testing framework to forecast results from future GW observations with improved observatories.
First laid out in~\cite{Carson_multiBandPRD} by the same authors, we discuss a simple method via a Fisher analysis discussed in Sec.~\nameref{sec:parameterized}.
To do so, one must first estimate the statistical uncertainties (manifesting from e.g. statistical detector noise) on $M_f$ and $\chi_f$ from the inspiral and merger-ringdown signals independently from the variance-covariance matrices $\bm{\Sigma}_{\I,\MR}$.
Next, we can estimate the systematic errors $\bm{\Delta_\mathrm{th}M_f}$ and $\bm{\Delta_\mathrm{th}\chi_f}$ potentially present in such an observation by following the work of Ref.~\cite{Cutler:2007mi}.
Such errors can manifest themselves from e.g. waveform mismodeling uncertainties by assuming a GR waveform template, and observing a non-GR signal.
In general, the systematic, or ``theoretical'', errors on parameters $\theta^a$ from such a scenario can be given as
\begin{equation}\label{eq:sys}
\Delta_\mathrm{th}\theta^a \approx \Sigma^{ab} \left( \lbrack \Delta A+iA_\GR\Delta\Psi \rbrack e^{i\Psi_\GR} \Big| \partial_b \tilde{h}_\GR \right),
\end{equation}
where $\Sigma^{ab} = (\Gamma^{-1})^{ab}$ is the covariance matrix found from the Fisher analysis, a summation over $b$ is implied, and $\Delta A \equiv A_\GR-A_\mathrm{non-GR}$ and $\Delta\Psi \equiv \Psi_\GR-\Psi_\mathrm{non-GR}$ are the differences between the amplitude and phase in GR and a given non-GR theory of gravity.

Combining both types of uncertainties found above, we can find the final probability distributions in the $M_f-\chi_f$ plane to be Gaussian
\begin{align}
P_{\I,\MR}\equiv & \frac{1}{2\pi\sqrt{|\bm{\Sigma}_{\I,\MR}|}}\exp \left\lbrack -\frac{1}{2} \left(  \bm{X} - \bm{X}^\GR_{\I,\MR} -\bm{\Delta_\mathrm{th}X}_{\I,\MR} \right)^\mathrm{T} \right. \nonumber \\
&\left. \times \bm{\Sigma}_{\I,\MR}^{-1}\left( \bm{X} - \bm{X}^\GR_{\I,\MR} -\bm{\Delta_\mathrm{th}X}_{\I,\MR} \right) \right\rbrack,\label{eq:pdf}
\end{align}
where $\bm{X}\equiv(M_f,\chi_f)$ contains the final state variables, $\bm{X}^\GR_{\I,\MR}$ contains their GR predictions from the inspiral and merger-ringdown portions respectively, and $\bm{\Sigma}_{\I,\MR}$ represents the covariance matrices of each portion.
By assuming a specific theory of gravity, one can simply compute increasing sizes of systematic uncertainties present in the above distributions corresponding to increasing magnitudes of non-GR effects.
While the statistical uncertainties manifest themselves as the \emph{size} of probability distributions, the systematic uncertainties will manifest as \emph{shifts} in the $(M_f,\chi_f)$ plane, ultimately increasing in magnitude until the inspiral and merger-ringdown signals no longer agree with one another.

Finally, once again the inspiral and merger-ringdown probability distributions can be transformed to the $(\epsilon,\sigma)$ coordinate system as discussed previously, allowing us to predict the magnitude of non-GR effects required to make such a distribution disagree with the GR value of $(0,0)$.
Once this point is reached, we can take the given size of non-GR effects to be a new constraint on that given theory.
The method described here can then be applied to alternative theories of gravity found in the literature, allowing us to forecast the power of the IMR consistency test in future GW observations for several theories of gravity.

\subsection{Current Status}\label{sec:IMRcurrent}

\begin{figure}
\begin{center}
\includegraphics[width=.8\columnwidth]{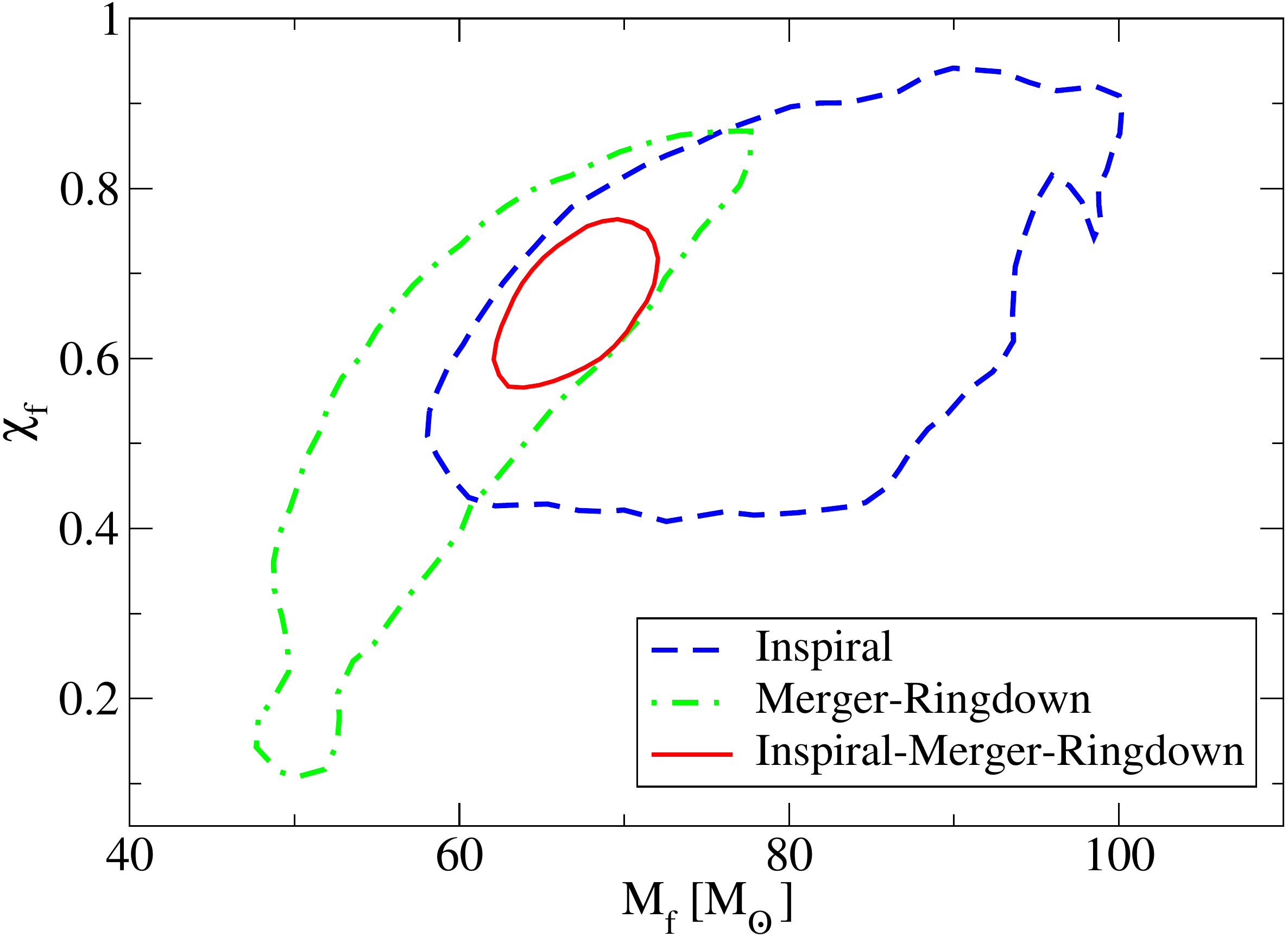}
\end{center}
\caption{90\% credible region contours of the inspiral, merger-ringdown, and complete inspiral-merger-ringdown waveform posterior distributions in the $M_f-\chi_f$ plane, for GW150914. The data was taken from~\cite{Abbott_IMRcon2}. 
}\label{fig:IMRDconsistency}
\end{figure}

Now let us discuss the current status of the IMR consistency tests in the GW community.
Figure~\ref{fig:IMRDconsistency} shows the posterior distribution of $M_f$ and $\chi_f$ from inspiral alone, merger-ringdown alone and the entire inspiral-merger-ringdown waveform for the event GW150914.
We display the results found by the LVC in~\cite{Abbott_IMRcon2} using a Bayesian analysis on the actual observed signal.
We see that, in the case of GW150914, the inspiral and merger-ringdown signals overlap significantly - showing very good agreement between the two  portions of the signal and the GR assumption is consistent with the data.

In Ref.~\cite{Abbott_IMRcon}, the LVC applied the IMR consistency test for the events GW150914, GW170104, GW170729, GW170809, GW170814, GW170818, and GW170823 as found in the GWTC-1 catalog~\cite{GW_Catalogue}.
The resulting posterior probability distributions from the IMR consistency test are displayed in Fig.~2 of~\cite{Abbott_IMRcon}.
All of the observed GW events considered in this test agree strongly with GR, indicating that if deviations from GR were to exist, we would need more sensitivity to discover them.
Further, the combined posterior probability distribution from all seven events similarly observes GR behavior.

Table~III of~\cite{Abbott_IMRcon} displays important quantities for each event considered, including the cutoff frequency between the inspiral and merger-ringdown signals, the SNR's within each region of the signal, and finally the GR quantile.
The latter quantity denotes the fraction of the posterior contained by an isoprobability contour passing directly through the GR value of $(0,0)$ in the $\epsilon$-$\sigma$ plane defined in Eq.~\eqref{eq:coordinates}.
Thus, smaller GR quantile fractions indicate better consistency between the observed signal and GR.
Such values occur between $7.8\%$ and $80.4\%$ for each of the events considered, with the largest value corresponding to GW170823, and the smallest from GW170814.
While the GR quantiles can vary by SNR (i.e. small-SNR signals will correspond to broad posteriors and smaller GR quantiles), we still see strong evidence in favor of GR as the true theory of gravity.

\subsection{Future Prospects}\label{sec:IMRfuture}

Now we will consider the future of IMR consistency tests with improved GW interferometers.
In Ref.~\cite{Ghosh_IMRcon}, the authors present the IMR consistency tests on ``golden'' BH binary coalescences.
Here, golden events occur as large-SNR events with total masses of $50\text{ M}_\odot-200\text{ M}_\odot$ that can be observed by ground-based GW observatories in all of the inspiral, merger, and ringdown portions of the signal.

The authors of~\cite{Ghosh_IMRcon} begin their analysis by simulating $\sim 100$ GR signals as observed by the advanced LIGO with its design sensitivity with SNRs of $25$.
This number of signals is potentially observable within one year of advanced LIGO operation according to several population synthesis models.
Figure~2 of the same paper displays the corresponding posterior probability distributions in the $(\Delta M_f/\overline{M}_f),\Delta \chi_f/\overline{\chi}_f)$ plane as computed via a Bayesian analysis.
All 100 posteriors agreed strongly with the GR value of $(0,0)$ and the combined posterior region could reach as low as a few percent.
This indicates the high degree of precision one can gain to test GR upon the successful combination of several GW observations.

Following this, the authors simulated general modified GR waveforms to predict the detectability of non-GR effects present within the signal with future observations.
To do this, the authors modified the GW flux at 2PN order by a constant factor of $\alpha_\text{modGR}$.
The authors then presented the IMR consistency test with $\alpha_\text{modGR}=400$ and found the IMR consistency test to be strongly inconsistent, as shown in Fig.~1 of~\cite{Ghosh_IMRcon}.
In particular, the inspiral and merger-ringdown posteriors clearly disagree with each other, and the transformed $(\Delta M_f/\overline{M}_f,\Delta \chi_f/\overline{\chi}_f)$ posterior clearly lies outside of the GR value of $(0,0)$.
In fact, the authors found that with such a signal, GR can be ruled out with a confidence of $\gg 99\%$, indicating the strong possibility with which one can detect deviations from GR with future observations.

\begin{figure}
\begin{center}
\includegraphics[width=.8\columnwidth]{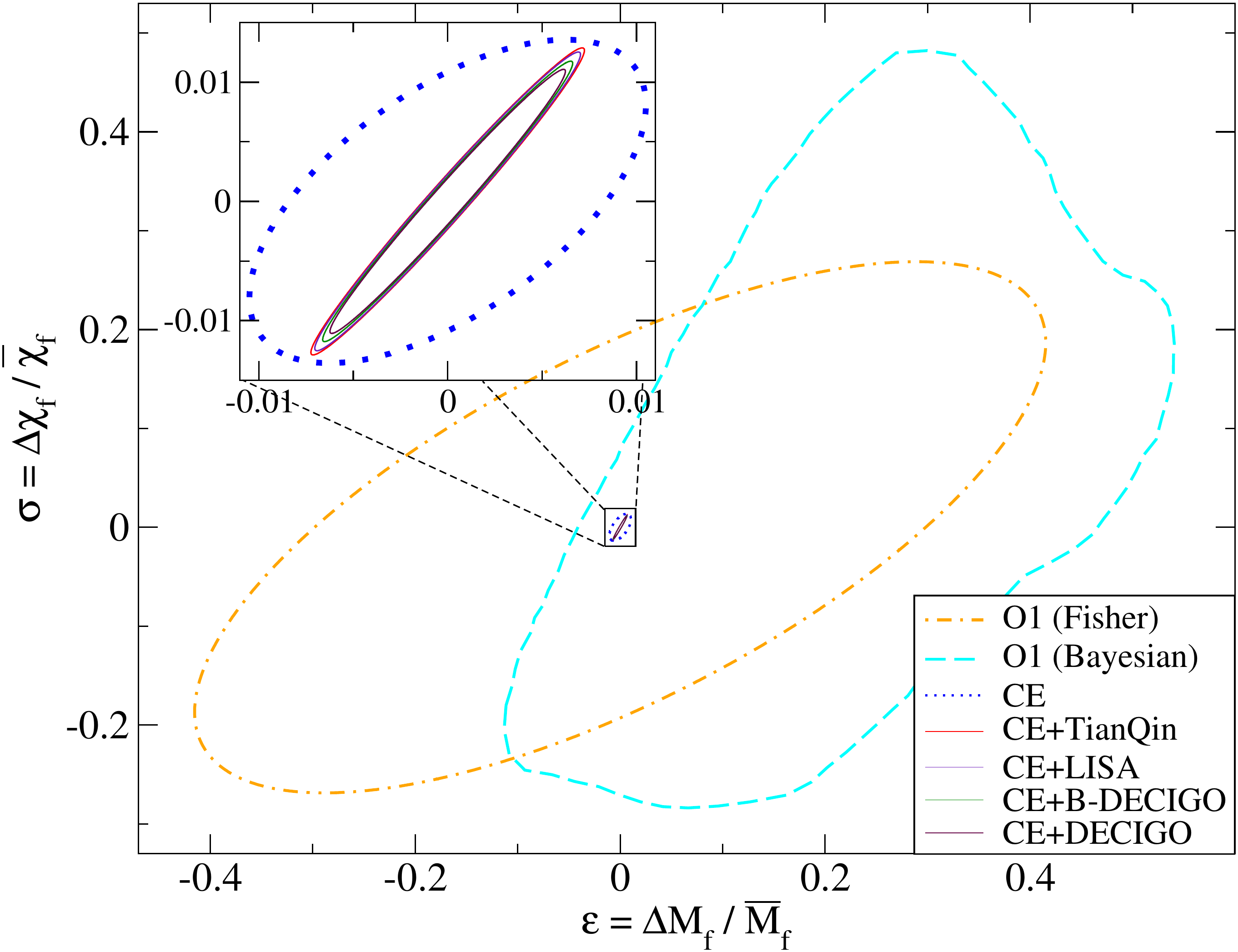}
\end{center}
\caption{
90\% credible region contours of the transformed probability distributions in the $\epsilon-\sigma$ plane (see Eqs.~\eqref{eq:coordinates}--\eqref{eq:transform}), describing the difference in the remnant mass and spin predictions between the inspiral and merger-ringdown estimate for GW150914-like event using the GR templates. 
Here we display the results for LIGO O1 (Fisher and Bayesian~\cite{Abbott_IMRcon} for comparison), CE, and the multi-band observations of CE and LISA, TianQin, B-DECIGO, and DECIGO. This figure is taken from Ref.~\cite{Carson_multiBandPRD}.
}\label{fig:IMRDconsistencyTransformed}
\end{figure}

Now let us switch our attention to future prospects of the IMR consistency test with third-generation ground-based detectors and space-based detectors. We show results using a predictive Fisher analysis as described in Secs.~\nameref{sec:Fisher} and~\ref{sec:IMRformalism}.
While this method is less robust than the ones discussed previously with Bayesian analyses, it offers one a quick method to obtain order-of-magnitude estimates on the future ability of the test in question.
By first estimating the intrinsic source parameters $m_1$, $m_2$, $\chi_1$, and $\chi_2$ using a Fisher analysis, the resulting Gaussian probability distributions are transformed into yet another two-dimensional Gaussian probability distribution in the $(M_f,\chi_f)$ plane as shown in Fig.~\ref{fig:IMRDconsistencyTransformed} for GW150914-like events with various future detectors. We also show the posterior distribution with the O2 noise curve obtained via a Fisher analysis, together with that obtained by LVC with a Bayesian analysis on the observed signal.
Here we see that the former predict reasonably-well those computed with the latter - both in the direction of correlation and size of the contour. 

Figure~\ref{fig:IMRDconsistencyTransformed} also presents the results for multi-band observations.
In particular, we considered the combination of future third-generation detector CE with space-based detectors LISA, TianQin, and (B-)DECIGO.
While the former detector can observe the merger-ringdown portion of the signal quite well, the latter detectors are apt at observing the early-inspiral regime of the signal.
The combination of the two was shown to decrease the area of the $(\Delta M_f/\overline{M}_f),\Delta \chi_f/\overline{\chi}_f)$ posterior by up to four-orders-of-magnitude as compared to that achievable by the first LIGO observing run shown in Fig.~\ref{fig:IMRDconsistency} (one-order-of-magnitude improvement from the ground-based-only detection by CE).
This considerable decrease in the area of the posterior points to the significant improvement in resolution one might gain with future observations.

\subsection{Applications to Specific Theories}

Although the IMR tests were originally designed to check the consistency of the GR assumption, it can be used to constrain specific theories (or spacetimes) as in the parameterized tests. We here review Refs.~\cite{Carson_QNM_PRL,Carson_BumpyQNM} by the same authors that made a first attempt in this direction.
In these investigations, we presented simple recipes one can use to introduce corrections to the inspiral, ringdown, and remnant BH mass and spin portions of the GW signal for various modified theories of gravity or spacetime metrics.
In this way, one can choose a specific theory of gravity or spacetime metric (and corresponding stress-energy tensor) and perform a theory-specific IMR consistency test of gravity with full-waveform corrections to obtain order-of-magnitude constraints on beyond-GR and beyond-Kerr parameters.

In particular, after applying the full-waveform corrections corresponding to the desired theory of gravity, one can estimate posterior probability distributions on the remnant BH mass and spin parameters using the Fisher analysis method.
Following this, one can slowly increase the magnitude of beyond-GR or beyond-Kerr corrections present in the hypothetical signal, calculating the corresponding systematic mismodeling uncertainties from Eq.~\eqref{eq:sys}.
Finally, one can compute the final posterior probability distribution via Eq.~\eqref{eq:pdf}.
Once the magnitude of beyond-GR effects are large enough to fail the IMR consistency test, we can take the corresponding value of the beyond-GR or beyond-Kerr parameter as a new constraint on that parameter if the the observed signal is consistent with GR.

\begin{figure}
\begin{center}
\includegraphics[width=.8\columnwidth]{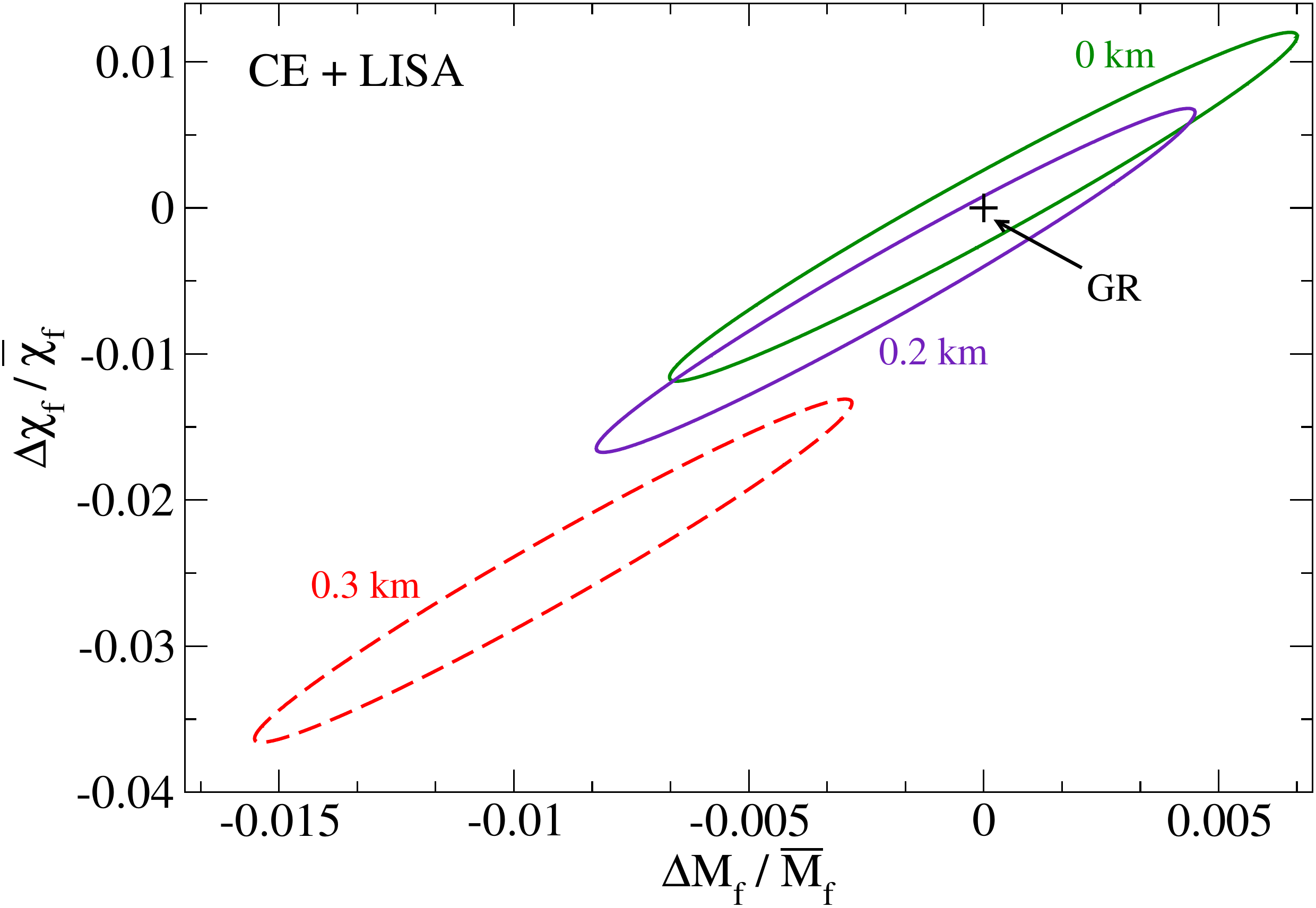}
\end{center}
\caption{
Future prospects for the IMR consistency to test EdGB gravity with multi-band observations. We assumed that both CE and LISA detects signals from a GW150914-like event. We show the 90\% credible posterior distribution for three difference choices of the EdGB coupling constant $\sqrt{\alpha_\EdGB}$. If the signal is consistent with GR, this test can place an upper bound of $\sqrt{\alpha_\EdGB} \lesssim 0.2$km, which is about 10 times stronger than the current bounds from low-mass X-ray binaries and GW observations. This figure is taken from~\cite{Carson_QNM_PRL}.
}\label{fig:IMR_EdGB}
\end{figure}

In~\cite{Carson_QNM_PRL,Carson_BumpyQNM}, the above technique was applied to both the EdGB gravity, and parameterized beyond-Kerr spacetime metrics proposed e.g. by Johannsen and Psaltis (JP)~\cite{Johannsen:2011dh}.
In the former case, constraints on the EdGB coupling parameter were shown to improve upon current constraints by up to an order-of-magnitude with future multi-band observations between ground-based and space-based detectors, CE and LISA (see Fig.~\ref{fig:IMR_EdGB}).
In the latter case, constraints on the JP deviation parameter via the space-based observation of extreme-mass-ratio-inspirals with LISA were found to be stronger than current bounds by up to three orders of magnitude.
The theory-specific constraints using the IMR consistency test were also found to be comparable with those found using the parameterized tests as discussed in Sec.~\nameref{sec:parameterized}.

\section{Gravitational-wave Propagation}

Up until now, we have been focusing on probing non-GR effects entering at the level of GW generation. Now, we switch gears to review probing non-GR effects during the GW propagation. 

\subsection{Modified Dispersion Relation}
\label{sec:mod_disp}

The GW propagation acquires non-GR corrections if the propagation speed of GWs is different from the speed of light. This is indeed the case in e.g. massive gravity theories in which the graviton has a non-vanishing mass~\cite{Will_mg}. A theory-agnostic way of probing such effects to the waveform was proposed by Mirshekhari \textit{et al}.~\cite{Mirshekari_MDR}, in which the generic modified dispersion relation of the graviton with energy $E$ and momentum $p$ has been proposed as 
\begin{equation}
\label{eq:disp_rel}
E^2 = (pc)^2 + \mathbb{A} (pc)^{\bar \alpha}\,.
\end{equation}
Here $\mathbb A$ is the overall magnitude of the correction to the dispersion relation while the index $\bar \alpha$ represents the $p$ dependence of the correction. Such a correction to the dispersion relation changes the gravitational waveform from the GR one. The correction can be mapped to the ppE formalism in Sec.~\ref{sec:parameterized} as~\cite{Mirshekari_MDR}
\begin{equation}
\beta = \frac{\pi^{2-\bar \alpha}}{1-\bar \alpha} \frac{D_{\bar \alpha}}{\lambda_{\mathbb{A}}^{2-\bar \alpha}} \frac{\mathcal M^{1-\bar \alpha}}{(1+z)^{1-\bar \alpha}}\,, \quad b = 3 (\bar \alpha-1)\,,
\end{equation}
where $\lambda_{\mathbb A} \equiv h \mathbb A^{1/(\bar \alpha-2)}$ is similar to a Compton wavelength while the distance $D_{\bar \alpha}$ is given by
\begin{equation}
D_{\bar \alpha} = \frac{(1+z)^{1-\bar \alpha}}{H_0} \int^z_0 \frac{(1+z')^{\bar \alpha-2}}{\sqrt{\Omega_m (1+z')^3 + \Omega_\Lambda}} dz'\,,
\end{equation}
where $z$ is the source redshift, $H_0$ is the current Hubble constant while $\Omega_m$ and $\Omega_\Lambda$ are the energy density of matter and dark energy respectively. The generic modification to the dispersion relation of the graviton in Eq.~\eqref{eq:disp_rel} can capture modifications in specific theories, including massive gravity, Doubly Special Relativity, extra dimension theories, Ho\v rava-Lifschitz gravity, multifractional spacetime theory and gravitational Standard Model Extension (see~\cite{Yunes_ModifiedPhysics} for details on the mapping).

\begin{figure}[h]
\begin{center}
\includegraphics[width=0.85\columnwidth]{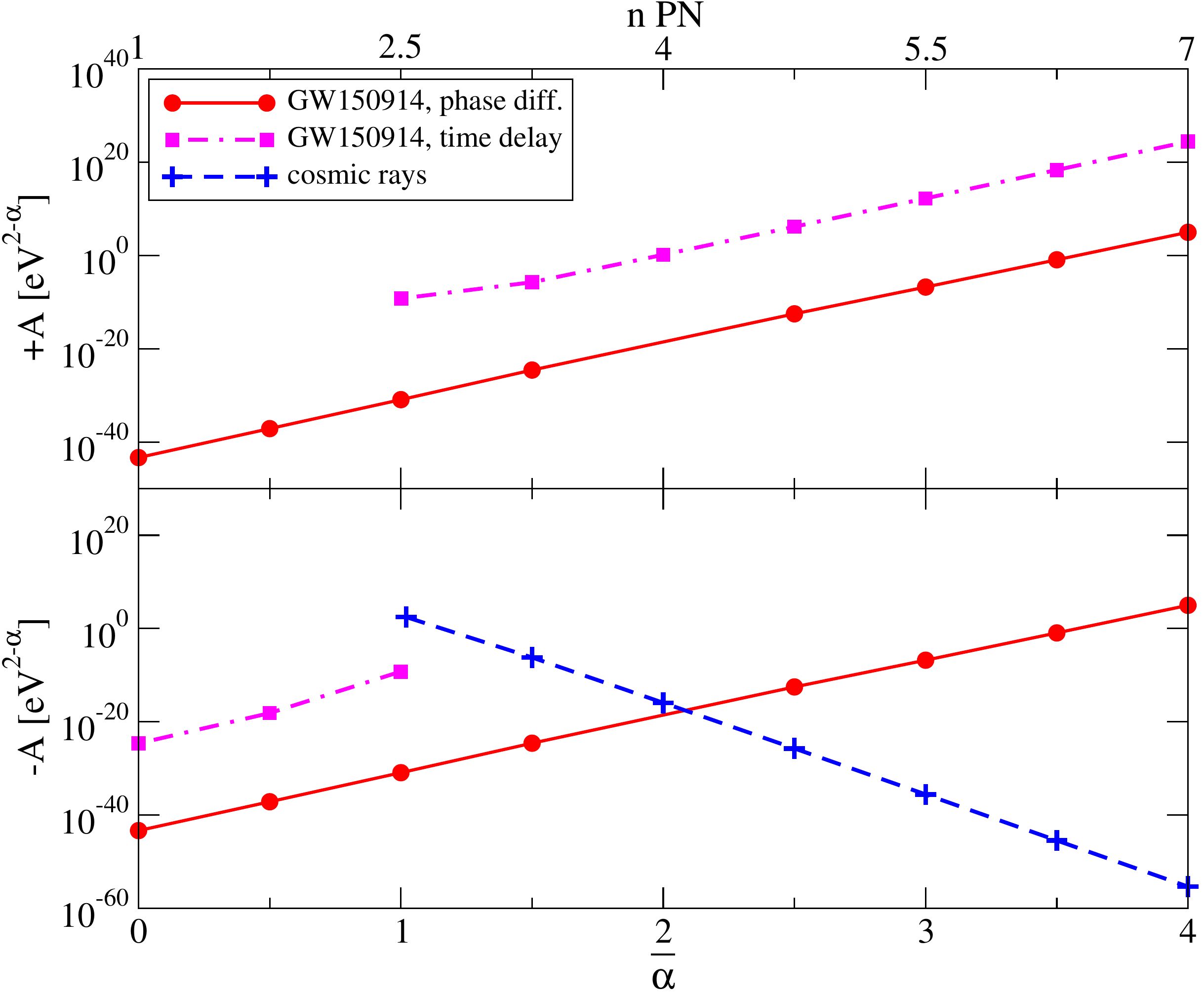}
\end{center}
\caption{Constraints on $\mathbb A$ in Eq.~(\ref{eq:disp_rel}) as a function of the index $\bar \alpha$ from GW150914. The top axis shows the corresponding PN order the correction enters to the phase. We show the one from the difference in the phase relative to GR (red solid)~\cite{Yunes_ModifiedPhysics} and that from the time delay between Hanford and Livingston detectors (magenta dotted-dashed)~\cite{Blas:2016qmn}. We also present the bound from cosmic rays (via the absence of gravitational Cherenkov radiation). Observe that GW150914 places unique bounds on the positive $\mathbb A$ parameter space.
This figure is taken and edited from~\cite{Yunes_ModifiedPhysics}.
}\label{fig:A}
\end{figure}

Figure~\ref{fig:A} presents the upper bound on $\mathbb A$ from GW150914 as a function of $\bar \alpha$~\cite{Yunes_ModifiedPhysics}. This bound was originally derived via a Fisher analysis, which was later confirmed by the LVC with a Bayesian analysis on the observed signal~\cite{Abbott_IMRcon}. We also present bounds from cosmic rays due to the absence of Cherenkov radiation, which can place stringent bounds on the subluminal propagation of GWs. Notice that the GW observations place a unique bound on the positive $\mathbb A$ parameter space. Figure~\ref{fig:A} also shows bound from GW150914 obtained by comparing the arrival time difference between the LIGO detectors at Hanford and Livingston~\cite{Blas:2016qmn}. Although such a bound is much weaker than that from the phase difference within the ppE framework, it can place a unique bound at $\bar \alpha = 2$ (corresponding to the correction to the GW propagation speed being independent of the GW frequency), where the phase correction degenerates with the time of coalescence. 

Future prospects on constraining $\mathbb A$ from GW150914-like events with upgraded detectors and multi-band observations are discussed in~\cite{Carson_multiBandPRD}. Since the corrections that we are considering enter at positive PN orders, ground-based detectors seem already good enough for probing such GW propagation effect with GW150914-like events and an addition of space-based detectors do not help much in this case.

\subsection{Graviton Mass and GW Propagation Speed}
\label{sec:mg}

When the graviton has a non-vanishing mass, there is a constant shift in the dispersion relation with a correction corresponding to $\bar \alpha = 0$. The bound on $\mathbb A$ from GW150914 at $\bar \alpha=0$ can be mapped to that on the graviton mass $m_g$ as $m_g < 10^{-22}$eV (see also Table~\ref{tab:summary}), which is consistent with what the LVC found~\cite{Abbott_IMRcon2}. The updated bound with all 10 binary BH events during the O1 and O2 runs combined is given by $m_g < 5 \times 10^{-23}$eV~\cite{Abbott_IMRcon}, which is slightly weaker than the updated Solar System bound~\cite{Will:2018gku}. 

We next discuss future prospects. The GW bound on $m_g$ is expected to improve significantly with GWs from supermassive BH binaries using LISA~\cite{Berti:Fisher,Chamberlain:2017fjl}. 
Future prospects for constraining $m_g$ with GW150914-like events including multi-band observations are studied in~\cite{Carson_multiBandPRD}. Multi-band observations can make the bound stronger than the single-band case by one order of magnitude, though such bounds are not as stringent as those from supermassive BH binaries with LISA.

The propagation speed of GWs, $c_\GW$, has also been constrained strongly from the binary NS event GW170817. Comparing the arrival time difference between GW and electromagnetic wave signals, the LVC placed the bound~\cite{Monitor:2017mdv} 
\begin{equation}
-3 \times 10^{-15} \leq \frac{c_\GW - c}{c} \leq 7 \times 10^{-16}\,.
\end{equation}
Such a stringent bound effectively ruled out many of modified theories of gravity, including various scalar-tensor and vector-tensor theories (see e.g.~\cite{Baker:2017hug}). We note that the comparison of GW and electromagnetic signals can also probe Shapiro time delays to constrain (or rule out) theories like relativistic MOdified Newtonian Dynamics (MOND) (see e.g.~\cite{Boran:2017rdn}), though some shortcomings have been pointed out in~\cite{Minazzoli:2019ugi}.

\subsection{Generic GW Propagation Tests}

A more generic way of testing GW propagation has been proposed by Nishizawa~\cite{Nishizawa:2017nef}. The author begins by taking a generic tensor perturbation $h_{ij}$ in cosmological background spacetime as~\cite{Saltas:2014dha}
\begin{equation}
\label{eq:tensor_pert}
h_{ij}'' + (2+\nu) \mathcal{H} h_{ij}'+(c_\GW^2 k^2 + a^2 m_g^2)h_{ij} = a^2 \Gamma \gamma_{ij}\,.
\end{equation}
Here $k$ is the wave number, $a$ is the scale factor, $\mathcal H = a'/a$ is the Hubble parameter in a conformal time, a prime represents a derivative with respect to the conformal time, $\nu$ is the Planck mass run rate and $\Gamma \gamma_{ij}$ is the source term due to anisotropic stress. The above equation reduces to GR when $c=1$ and $(\nu,m_g,\Gamma) = (0,0,0)$. The generic modified dispersion relation considered in Sec.~\ref{sec:mod_disp} corresponds to $c_\GW = c_\GW(k)$ and $(\nu,\Gamma)=0$. Corrections to the waveform from the non-GR parameters in Eq.~\eqref{eq:tensor_pert} can be mapped to the ppE (and gIMR) formalism~\cite{Nishizawa:2017nef}. Future prospects of probing the generic GW propagation with second-generation ground-based detectors are discussed in~\cite{Nishizawa:2017nef}.

Let us now focus on the case where $\nu \neq 0$.  The parameter $\nu$ acts as a friction term that changes the amplitude, and if it is independent of $k$, the corresponding ppE parameter is given by
\begin{equation}
\alpha = - \frac{1}{2} \int^z_0 \frac{\nu}{1+z'} dz'\,, \quad a=0\,.
\end{equation}
This effect can be absorbed into the luminosity distance $d_L$. This means that the luminosity distance $d_L^\GW$ measured with GWs would be different from $d_L^\EM$. Such a difference can generically be parameterized with two parameters $\Xi_0$ and $n$ as~\cite{Belgacem:2018lbp,Belgacem:2019pkk}
\begin{equation}
\Xi(z) \equiv \frac{d_L^\GW(z)}{d_L^\EM(z)} = \Xi_0 + \frac{1-\Xi_0}{(1+z)^n}\,,
\end{equation}
which is related to $\nu$ as
\begin{equation}
\nu(z) = - \frac{d \ln \Xi(z)}{d \ln(1+z)} = \frac{n(1-\Xi_0)}{1-\Xi_0+\Xi_0(1+z)^n}\,.
\end{equation}
The mapping between $(\Xi_0,n)$ and parameters in various example theories is presented in Table~1 of~\cite{Belgacem:2019pkk}. Future prospects for measuring and constraining $\Xi_0$ with GW standard sirens using third-generation ground-based detectors and LISA are discussed in~\cite{Belgacem:2018lbp} and~\cite{Belgacem:2019pkk} respectively.

\subsection{Amplitude Birefringence in Parity Violation}

One way to probe the gravitational parity violation is through the amplitude birefringence~\cite{Alexander:2007kv,Yagi:2017zhb}. When GWs propagate, the amplitude of right-handed and left-handed circular polarizations behave differently. 

Let us take Chern-Simons (CS) gravity as an example.
We begin by considering a metric perturbation under the Friedmann-Robertson-Walker spacetime given by
\begin{equation}
ds^2 = a^2(\eta) \left[ -d\eta^2 +(\delta_{ij} + \bar h_{ij}) d\chi^i d\chi^j \right]\,,
\end{equation}
where $\chi^i$ is the comoving spatial coordinates, while $\bar h_{ij}(\eta,\chi^i)$ is the comoving metric perturbation. We decompose this metric perturbation into right-handed ($\bar h_R$) and left-handed ($\bar h_L$) circular polarizations, which are related to the plus and cross mode polarizations as $\bar h_{R} = (\bar h_+ - i \bar h_{\times})/\sqrt{2}$ and  $\bar h_{L} = (\bar h_+ + i \bar h_{\times})/\sqrt{2}$, respectively.
We further decompose $\bar h_{R,L}$ into the amplitude $\mathcal{A}_{R,L}$ and phase $\phi(\eta)$ as
\begin{equation}
\bar h_{R,L} = \mathcal{A}_{R,L} e^{-i\left[ \phi(\eta)-\kappa n_k \chi^k \right]}\,,
\end{equation}
where $\kappa$ is the conformal wave number while $n^k$ represents the unit vector pointing towards the direction of GW propagation.
Plugging these into the modified field equations, one finds an equation for the phase (dispersion relation) given by
\begin{equation}
\label{eq:DR_CS}
i \phi'' + (\phi')^2 - \kappa^2 = -2i \frac{\mathcal S_{R,L}'}{\mathcal S_{R,L}} \phi'\,, \quad
\mathcal S_{R,L} \equiv a \sqrt{1-\lambda_{R,L} \frac{\kappa \vartheta'}{a^2}}\,,
\end{equation}
with $\lambda_{R}=+1$, $\lambda_L=-1$ and $\vartheta$ representing the scalar field in CS gravity.

We next impose the following assumptions: $(\phi')^2 \gg \phi''$,  
$\kappa \gg \mathcal{S}'_{R,L}/\mathcal{S}_{R,L}$.
The second condition is satisfied when $\kappa \gg \mathcal H$ and imposing the weak CS approximation given by
\begin{equation}
\label{eq:weak_CS}
\kappa |\vartheta'| \ll a^2\,, \quad \kappa |\vartheta'' | \ll 2 a^2 \mathcal{H}\,.
\end{equation}
Under these assumptions, one can solve the dispersion relation in Eq.~\eqref{eq:DR_CS}. Evaluating the solution at the current conformal time of $\eta = 1$, one finds
\begin{equation}
\label{eq:phi_RL}
\phi_{R,L}(1) = \pm \kappa (1-\eta_s)+i \lambda_{R,L} \pi f \dot \Theta\,.
\end{equation}
Here $\eta_s$ is the conformal time at which GWs are emitted while $\dot \Theta \equiv  \dot \vartheta_0 -(1+z) \dot \vartheta_s$, 
where the subscripts 0 and $s$ represent a quantity being evaluated at $\eta=1$ and $\eta=\eta_s$ respectively. An over-dot refers to a derivative with respect to the physical time $t$. 

Notice that the CS correction in Eq.~\eqref{eq:phi_RL} is purely imaginary, which means that such a correction actually affects the amplitude. Moreover, it is proportional to $\lambda_{R,L}$, which leads to the conclusion that if one of the circular polarization is amplified, the other one is suppressed. This effect can be summarized as
\begin{equation}
\bar h_{R,L} = \bar h_{R,L}^\GR \left( 1+\lambda_{R,L} \pi f \dot \Theta \right)\,,
\end{equation}
where $\bar h_{R,L}^\GR$ is the comoving metric perturbation for circular polarizations in GR. This can be mapped to the PPE framework and the amplitude correction enters at 1.5PN order relative to GR.

Let us now discuss the current bounds and future prospects of probing parity violation from GW amplitude birefringence. We begin with isolated GW sources. Unfortunately, current observation of GW150914 cannot place any meaningful bounds because they do not satisfy the weak CS bound in Eq.~\eqref{eq:weak_CS}~\cite{Yagi:2017zhb}. On the other hand, if aLIGO with its design sensitivity or CE detects signals from a GW150914-like event, one should be able to place bounds on parity violation from birefringence. Another interesting possibility of using multi-messenger observations to probe amplitude birefringence is discussed in~\cite{Yunes:2010yf}.

\begin{figure}[h]
\begin{center}
\includegraphics[width=0.8\columnwidth]{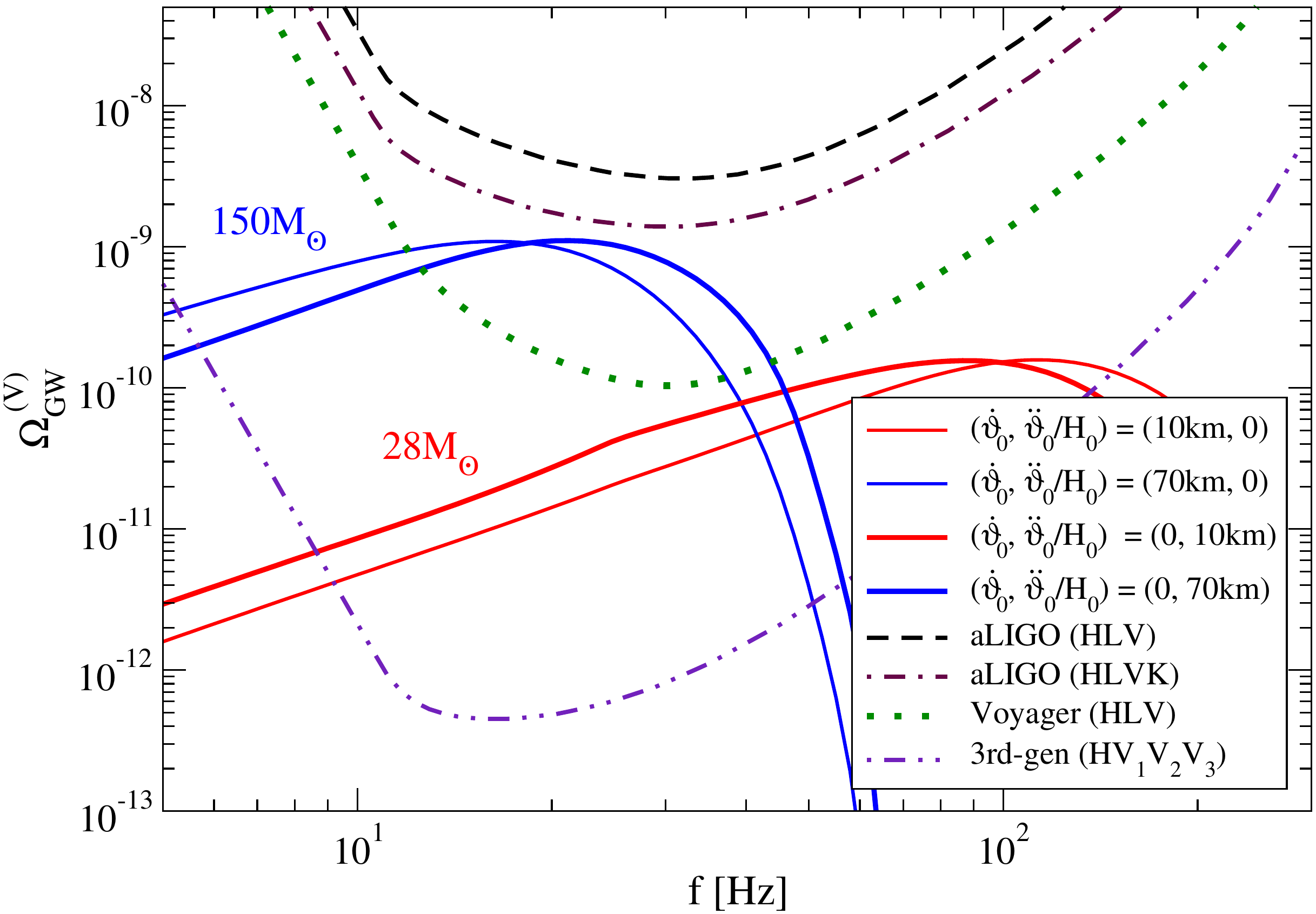}
\end{center}
\caption{Astrophysical GW background from binary BH with two different average chirp masses for the Stokes' V-mode (parity-violating mode) in dCS gravity, together with the power-law integrated sensitivity curves for various combinations of detectors at Hanford (H), Livingston (L) Virgo (V) and KAGRA (K) sites. For the spectrum, we consider different combinations of $\dot \vartheta_0$ and $\ddot \vartheta_0/H_0$. If the spectrum goes above the sensitivity curve at any frequency, the signal-to-noise ratio is beyond unity. Parity violation in the V-mode GW background may be detected with thrid-generation detectors. This figure is taken from~\cite{Yagi:2017zhb}.
}\label{fig:GWB}
\end{figure}

We next look at using stochastic GW background (GWB) from binaries of stellar-mass BHs, which we assume to be stationary, Gaussian and isotropic. The quadratic expectation values for each polarization is given by
\begin{equation}
\begin{pmatrix}
\langle \tilde h_R(f,n)\, \tilde h_R^*(f',n') \rangle \\
\langle \tilde h_L(f,n)\, \tilde h_L^*(f',n') \rangle 
\end{pmatrix}
=\frac{1}{2} \delta (f-f') \delta^2(n , n') \times
\begin{pmatrix}
I(f) + V(f) \\
I(f) - V(f)
\end{pmatrix}\,,
\end{equation}
where the angular brackets denote ensemble averaging while $I(f)$ and $V(f)$ are the Stokes parameters corresponding to the total intensity and parity-violation respectively. We will focus on the latter, which is absent in the standard GR spectrum of GWs so that its detection will hint at the presence of parity violation. The fractional energy density spectrum of the V-mode is given by 
\begin{equation}
\Omega_\GW^{(V)} (f) \equiv  \frac{4\pi^2 f^3}{\rho_c} V(f)\,,
\end{equation}
where $\rho_c$ is the critical energy density of the Universe.

Figure~\ref{fig:GWB} presents the spectrums of the V-mode for various choices of the evolution of the scalar field and the average chirp mass of binary BHs. We also show the power-law integrated sensitivity curves (with an SNR of 1) for networks of 2nd and 3rd generation ground-based detectors. Observe that such spectrums may be detectable with the 3rd-generation detectors.

Amplitude birefringence for a single tensor perturbation can be extended to chiral mixing between two tensor perturbations. Such a chiral mixing is one example of  GW oscillations (similar to neutrino oscillations) considered in~\cite{Jimenez:2019lrk} which describes the mixed flavor of two tensor perturbations. Other examples include  friction, velocity and mass mixing. 
The flavor mixing between the tensor perturbations $h^{(1)}_{ij}$ and $h^{(2)}_{ij}$ can be written in a unified framework as
\begin{equation}
\left( \frac{d^2}{d\eta^2} + \hat \nu + \hat C k^2 + \hat N k + \hat M \right) 
\begin{pmatrix}
h^{(1)} \\
h^{(2)}
\end{pmatrix}=0\,.
\end{equation}
Here, $\hat \nu$ is the friction matrix, $\hat C$ is the velocity matrix, $\hat N$ is the chirality matrix,  $\hat M$ is the mass matrix, and $k$ is the wave number. It would be interesting to study how accurately one can probe each of these GW oscillation effects with current and future GW observations.

\section{Open Questions}

We end this chapter by presenting several open questions that need to be addressed to further improve the ability of testing GR with GWs:

\begin{enumerate}
\item \emph{Higher PN corrections in the inspiral}: One of the current major issues is that methods for extracting physical implications from theory-agnostic tests are limited due to the lack of complete inspiral-merger-ringdown waveforms in theories beyond GR. In the inspiral part, most studies focused on finding the leading PN corrections to the waveform. Higher PN corrections that are usually neglected are necessary to construct complete waveforms in non-GR theories, especially since the PN approximations become inaccurate close to merger. Towards constructing complete waveforms, one needs to consider generic binary systems, including spin precession and large eccentricity.
\item \emph{Corrections in the merger phase}: Numerical relativity simulations of compact binary mergers have been carried out in a few non-GR theories, such as scalar-tensor theories and EdGB/dCS gravity~\cite{Witek:2018dmd,Okounkova:2019dfo}, though more simulations need to be performed with a broader class of theories, with parameters (mass ratio, spins, etc.) varied systematically. When carrying out numerical relativity simulations, one also needs to check the well-posedness of the theories, though one way to overcome this is to treat the theories as effective theories and solve order by order in theoretical parameters that are assumed to be small, as done within EdGB/dCS simulations~\cite{Witek:2018dmd,Okounkova:2019dfo}.
\item \emph{Corrections in the ringdown phase}: From numerical relativity simulations, one can also find corrections in the ringdown phase. Another approach is to find corrections to the QNMs through the BH perturbation. However, most studies focused on non-rotating BHs, and we still lack a systematic analysis of including spins. Perhaps the first step is to carry out calculations for slowly-rotating BHs. Ideally, one would want to extend the analysis for arbitrary spins, though Kerr-like solutions are still lacking in many modified theories of gravity. 
\item \emph{Phenomenological waveforms}: Once the above inspiral, merger and ringdown corrections to the waveform are computed, one can then attempt to construct phenomenological IMR waveforms in specific non-GR theories. Such waveforms will be useful for e.g. extracting physics from merger-ringdown gIMR parameters. It would be also interesting to investigate whether one can construct phenomenological waveforms for scalar and vector polarization modes that are typically present in non-GR theories.
\item \emph{GW memory}: There are nonlinear effects in the waveform that are subdominant and have not studied much in theories beyond GR. One example is the GW memory that may be measured with future ground-based and space-based detectors. It would be important to extend the Bondi-Sachs formalism and compute the memory waveforms in modified theories of gravity, including scalar and vector memories. Such nonlinear effects may give new insights to beyond-GR theories.
\item \emph{Cosmological screening}: Another nonlinear effects that may arise in non-GR theories motivated to explain cosmological problems are various screening mechanisms~\cite{Jain:2010ka}. Some studies exist on how such mechanisms affect the GW generation~\cite{deRham:2012fw} and propagation~\cite{Perkins:2018tir}, though more analysis needs to be done along this direction.
\item \emph{Astrophysical systematics}: In order to test GR, one needs to have systematic errors under control. One possible source of systematics is through astrophysical effects~\cite{Barausse:2014tra}, such as accretion disks~\cite{Kocsis:2011dr}, tidal resonances~\cite{Bonga:2019ycj} and dark matter halos. More work needs to be done on how such systematics may limit the ability of testing GR with GWs.
\item \emph{Electromagnetic counterparts}: GW170817 opened a new window for multi-messenger astronomy, and we expect to find similar events in the future. It would be interesting to perform merger simulations of binary NSs in modified theories of gravity and predict how the electromagnetic counterpart signals may get modified from GR. One can then compare the GW and electromagnetic signals to perform multi-messenger tests of GR, which is beyond what has already been done by e.g. comparing the arrival time difference between gravitons and photons to constrain the propagation speed of GWs and the graviton mass~\cite{Monitor:2017mdv}. 
\end{enumerate}



\section*{Acknowledgements}
Z.C. and K.Y. acknowledge support from the Owens Family Foundation.
K.Y also acknowledges support from NSF Award PHY-1806776, NASA Grant 80NSSC20K0523 and a Sloan Foundation Research Fellowship.  

\bibliography{Zack.bib}

\end{document}